\documentclass[twocolumn]{aastex631}

\usepackage{physics}
\usepackage{enumitem}

\usepackage{needspace}
\usepackage{etoolbox}

\preto\section{\Needspace{6\baselineskip}}
\preto\subsection{\Needspace{5\baselineskip}}

\hypersetup{
  colorlinks=true,
  linkcolor=blue,
  citecolor=blue,
  urlcolor=blue
}

\newcommand{\mearth}{M$_{\oplus}$\,}

\begin{document}

\title{Sub-Neptune Memories I: Implications of Inefficient Mantle Cooling and Silicate Rain}


\correspondingauthor{Roberto Tejada Arevalo}
\email{arevalo@princeton.edu}

\author[0000-0001-6708-3427]{Roberto Tejada Arevalo}
\affiliation{Department of Astrophysical Sciences, Princeton University, 4 Ivy Lane,
Princeton, NJ 08544, USA}
\author[0000-0002-2006-7769]{Akash Gupta}
\affiliation{Department of Astrophysical Sciences, Princeton University, 4 Ivy Lane,
Princeton, NJ 08544, USA}
\affiliation{Department of Geosciences, Princeton University, 11 Ivy Lane, Princeton, NJ 08550, USA}

\author[0000-0002-3099-5024]{Adam Burrows}
\affiliation{Department of Astrophysical Sciences, Princeton University, 4 Ivy Lane,
Princeton, NJ 08544, USA}

\author[0009-0003-0497-4961]{Donghao Zheng}
\affiliation{Department of Geosciences, Princeton University, 11 Ivy Lane, Princeton, NJ 08550, USA}

\author[0000-0003-3980-7808]{Yao Tang}
\affiliation{Department of Astronomy and Astrophysics, University of California, Santa Cruz, 1156 High Street, Santa Cruz, CA 95064, USA}

\author[0000-0001-5441-2797]{Jie Deng}
\affiliation{Department of Geosciences, Princeton University, 11 Ivy Lane, Princeton, NJ 08550, USA}

\begin{abstract}

We explore the evolution of sub-Neptune (radii between $\sim$1.5 and 4 R$_\oplus$) exoplanet interior structures using our upgraded evolution code, \texttt{APPLE}, which self-consistently couples the thermal and compositional evolution of the whole structure. We incorporate stably stratified regions with convective mixing and, for the first time, ab initio results on the phase separation of silicate-hydrogen mixtures to model silicate rain in sub-Neptune envelopes. We demonstrate that inefficient mantle cooling can retain sufficient heat to Gyr ages: inefficient heat transport from mantle to envelope alone keeps radii $\sim$10\% larger than predicted by adiabatic models at late times. Silicate rain can contribute an additional $\sim$5\% to the radius, depending on envelope mass and initial metal abundance. The silicate-hydrogen immiscibility region may lie in the middle or even upper envelope, far above the envelope-mantle boundary layer, and bifurcates the envelope into two an upper, hydrogen-rich region and a lower, metal-rich region above the mantle. If silicate rain occurs, atmospheres should appear depleted of silicates while radii remain inflated at late ages. To demonstrate the effects of inefficient mantle cooling, we present interior evolution models for GJ 1214 b, K2-18 b, TOI-270 d, and TOI-1801 b, showing that hot, liquid silicate mantles with thin envelopes reproduce their radii and mean densities, providing an alternative to water-world interpretations. These results imply that bulk compositions inferred from mean density must account for the mantle thermal state and the envelope mixing/phase-separation history; such thermal ``memories'' may constrain formation entropies and temperatures when metallicities are more precisely measured.

\end{abstract}


\section{Introduction}

Sub-Neptune exoplanets, typically defined as having radii between $\sim$1.5 and 4 R$_\oplus$ and masses between 3 and 15 M$_\oplus$, constitute a substantial fraction of the close-in planet population \citep[e.g.,][]{Howard2012, Fulton2017, Zhu2018, Petigura2018, WinnPetigura2024}. Indeed, close-in planets in this radius range appear to be an order of magnitude more common than those between 4 and 16 R$_\oplus$ \citep{Petigura2018}. Because of their ubiquity, sub-Neptunes can provide critical constraints on planet formation and evolution \citep{Bean2021, Parc2024}. While bulk density measurements derived from transit and radial velocity surveys suggest that some sub-Neptunes are consistent with rocky compositions, others require a volatile envelope with a likely hydrogen-helium (H-He) layer enriched with heavy elements \citep[e.g.,][]{Rogers2015, Wolfgang2016}. Some of these densities appear to be too low for predominantly rocky compositions, leading to the proposal of ``water worlds'' as a distinct class \citep{Madhusudhan2021, Luque2022}. However, because inferences on bulk composition usually rely on mass and radius alone, they are subject to significant degeneracies. These include uncertainties regarding internal thermal states \citep{Vazan2017, Vazan2018a}, envelope and atmosphere metallicities \citep{Lopez2014, OwenWu2017, Gupta2019}, phase separation \citep[e.g.,][]{Chachan2018, StixrudeGilmore2025, Rogers2025}, and convective inhibition induced by compositional gradients \citep{Misener2022, Vazan2024}.

In many sub-Neptune evolution models \citep[e.g.,][]{Lopez2012,Owen2013, ChenRogers2016, Lopez2014, OwenWu2017, Gupta2019,Gupta2020a,Gupta2022a}, the cooling of the mantle and core is assumed to follow the cooling rate of the envelope, implying that the total energy budget of the deep interior is inconsequential to the cooling rate. However, the initial energy content of sub-Neptunes may reside primarily in their mantles and cores, which may not cool at the same rate as their envelopes \citep{Vazan2017}. Contrary to the assumption that mantle heat dissipates on $\sim$100 Myr timescales, \cite{Vazan2017} found that initial mantle temperatures can alter the planetary radius by $\sim$15\% and that magma oceans may persist into late ages \citep{Vazan2018a,Tang2025}. \cite{Tang2025} also found that sub-Neptune mantles can continue to be liquid well into late evolutionary stages due to high instellation and their relatively thick envelopes. Moreover, \cite{Eberlein2025} demonstrated that when non-adiabatic, inhomogeneous structures are considered, interior conductivities can affect radius evolution by up to $\sim$25\%. These results indicate that if sub-Neptune mantles cool inefficiently, planets may retain inflated radii at Gyr ages, biasing inferences about average (or ``bulk'') composition. Capturing these effects requires a treatment of heat transport comparable to that used in stellar evolution, yet most current sub-Neptune models lack this capability. 

In this work, we explore the possibility that average densities currently attributed to water worlds can also be explained by hot, inefficiently cooling mantles, and the effects of convectively stable layers in thin ($\sim$5\% by mass) sub-Neptune envelopes. We present state-of-the-art sub-Neptune interior evolution models using our \texttt{APPLE} planet evolution code \citep{Sur2024a}, which we upgrade here to model sub-Neptune interior evolution. We build on the work of \cite{Vazan2017}, \cite{Vazan2018a}, more recent sub-Neptune evolution models from \cite{Tang2025}, and recent advances in Solar System giant modeling \citep{Tejada2025, Sur2025a, Tejada2025b}. We focus specifically on the role of the initial thermal state of the mantles and cores, the impacts of coupled envelope-mantle-core cooling, and silicate phase separation in the envelope. To accurately model these processes, we implement a modified mixing-length theory \citep[MLT;][]{Bohmvitense1958MLT, Sasaki1986, Kippenhahn1990} in the mantle and core that incorporates latent heat and time-dependent radiogenic heating. Furthermore, we deploy updated conductivities, opacities, and equations of state for the envelope, alongside atmospheric models that account for stellar irradiation.

Our methodology, including equations of state (EOSes), microphysics, code upgrades, and silicate rain prescription, is described in Section~\ref{sec:methods}. In Section~\ref{sec:evolution}, we present the thermal and radius evolution of our models as a function of mass and initial mantle temperature, the effects of stably-stratified layers, and we model the effects of silicate rain as a function of mass, equilibrium temperature, and envelope mass. We also present case studies for GJ 1214 b \citep{Charbonneau2009}, K2-18 b \citep{Monet2015}, TOI-270 d \citep{VanEylen2021}, and TOI-1801 b \citep{Mallorquin2023}. We discuss our results and their implications in Section~\ref{sec:discussion} and offer a summary of our findings and concluding remarks in Section~\ref{sec:conclusion}.

\section{Methods}\label{sec:methods}

\texttt{APPLE} solves the equations of mass, momentum, energy, and species conservation using the standard set of stellar structure equations, as discussed in Section 2 of \cite{Sur2024a}. We deploy the Henyey relaxation method \citep{Henyey1964}, which solves the structure and evolution equations in mass coordinates rather than radial coordinates, as described in Section 8 of \cite{Sur2024a}. \texttt{APPLE}'s design is inspired by stellar evolution codes, such as \texttt{MESA} \citep{Paxton2011, Paxton2013, Paxton2018} and entropy convective criterion notation \citep[e.g.,][]{Lattimer1981}. \texttt{APPLE} is built with the flexibility to incorporate various equations of state, such as those of hydrogen, helium, water, and silicates, and their mixtures \citep{Tejada2024}, at and for any thermal and compositional state. This gives \texttt{APPLE} the flexibility to impose any initial structure and carry out its thermal-compositional evolution, as already deployed to model the inhomogeneous and non-adiabatic evolution of Jupiter and Saturn \citep{Tejada2025, Sur2025a} and Uranus and Neptune \citep{Tejada2025}. Such features used in this work are described in the following sub-sections. 

\subsection{EOS}\label{subsec:eos}

\begin{figure}[ht!]
\centering
\includegraphics[width=0.43\textwidth]{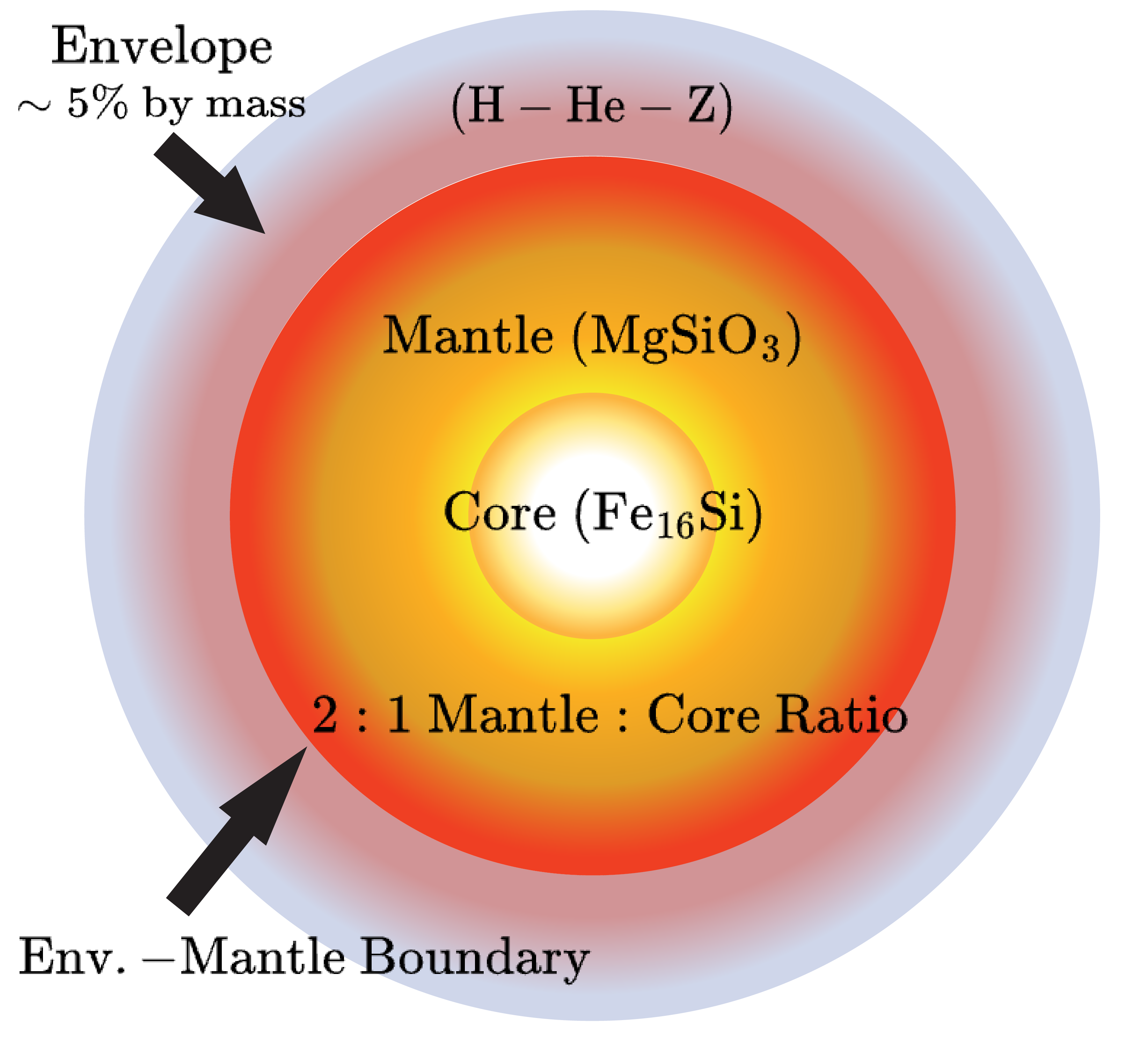}
\caption{Interior structure schematic for the interior composition of sub-Neptune models presented in this work. Color is an approximate indicator of temperature (specific values depend on the model). The envelope ($\sim$5\% by mass) is a mixture of hydrogen, helium, and heavy elements ($Z$) represented by either water (Section~\ref{subsec:entropy}) or silicates (Sections~\ref{subsec:stratification}, \ref{subsec:miscibility}). The mantle composition is MgSiO$_3$, and the core composition is Fe$_{16}$Si. We maintain a 2:1 mass ratio between the mantle and the core. For this work, the EMB represents a steep compositional gradient and thus controls mantle cooling by transferring heat to the envelope via conduction. Since conductive flux alone is insufficient to transport the entire interior heat to the envelope, a steep temperature gradient forms at the EMB, as seen in all models in this work.}
\label{fig:fig1}
\end{figure}

All structures in this work are divided into an envelope composed of hydrogen and helium mixed with varying amounts of water or silicates (described below), a silicate rocky mantle, and an iron-rich core. The illustration in Figure~\ref{fig:fig1} summarizes the interior composition structure assumed in this work. The metals in the envelope are represented by the pure water ``AQUA'' EOS of \citep{Haldemann2020} in Sections \ref{subsec:entropy} and \ref{subsec:case_studies}, a mixture of pure water and silicates in Section \ref{subsec:stratification}, and pure silicates in Section \ref{subsec:miscibility}. The AQUA EOS incorporates the water EOS of \cite{Mazevet2019}, which has an entropy error corrected by \cite{Mazevet2021}. We use this corrected version here, although such a correction is inconsequential \citep[see Appendix A of][]{Tejada2025b} and does not affect the density. While it is unlikely that water and silicates are the only species in sub-Neptune envelopes, this simplification is common to all evolutionary models (even those of the Solar System planets). Other species, such as methane and ammonia, could also be prevalent, as they likely are on Uranus and Neptune \citep{Nettelmann2016, Bethkenhagen2017, Militzer2025, Tejada2025b}.

Throughout this work, sub-Neptune mantles are characterized by pure MgSiO$_3$. We use an updated EOS for liquid MgSiO$_3$ calculated by \cite{Luo2025}, derived from ab initio methods at high pressures ($\sim$1,200 GPa) and temperatures ($\sim$14,000 K). This liquid MgSiO$_3$ EOS includes calculations of the specific entropy using the thermodynamic integration technique \citep[e.g., as applied in][for the H-He EOS]{Militzer2013}. For the solid phase of MgSiO$_3$, we use the post-perovskite third-order Birch-Murnaghan (BM3) EOS with thermal components of \cite{Sakai2016} for pressures $\geq 95$ GPa. For pressure ranges $23 \leq P < 95$, we use the solid perovskite EOS of \cite{Tange2012}. The exact perovskite/post-perovskite transition can occur at pressures greater than 95 GPa, as informed by the temperature dependence calculated by \cite{Deng2023}. For pressures $<23$ GPa, we use the enstatite BM3 EOS parameters reported in \cite{Angel2002}. Throughout this work, we use the MgSiO$_3$ melting curve of \cite{Fei2021} to decide the liquid-solid phase transition of the mantle.

\begin{figure}[ht!]
\centering
\includegraphics[width=0.43\textwidth]{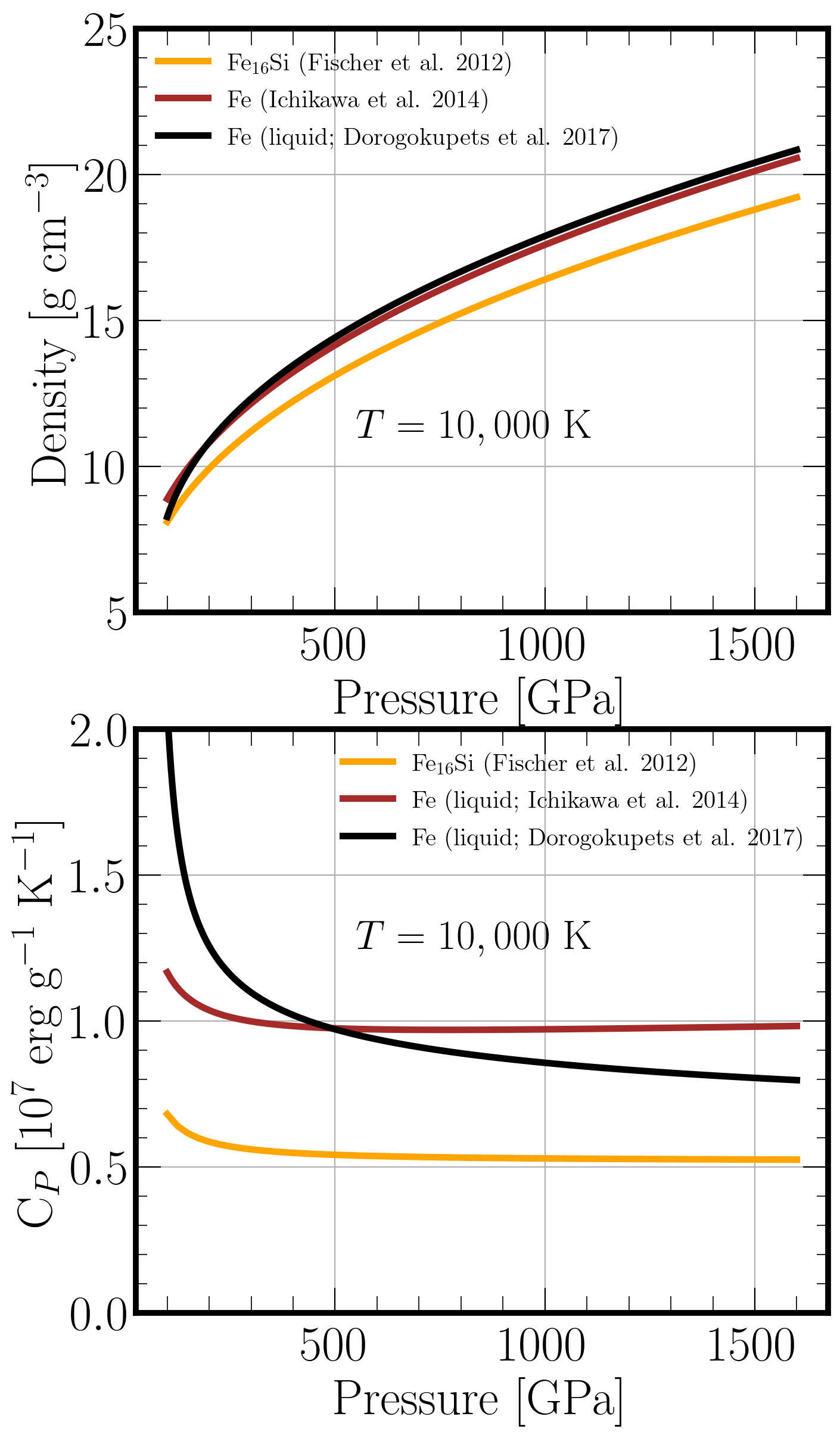}
\caption{Example isotherms of Fe$_{16}$Si EOS from \cite{Fischer2012}, and the liquid iron EOSes of \cite{Ichikawa2014} and \cite{Dorogokupets2017Thermodynamics6000K},  at a temperature of 10,000 K. The density differences (top panel) above 200 GPa range between 7\% and 10\%, and the isobaric heat capacity, C$_P$ (bottom panel), differs by nearly 50\%. Present models of Earth's interior indicate that its liquid outer core may be composed of iron alloys. For simplicity, we apply the Fe$_{16}$Si EOS of \cite{Fischer2012} in the cores throughout this work. }
\label{fig:fig2}
\end{figure}

 We assume that the core is well represented by the iron alloy, Fe$_{16}$Si EOS, as calculated by \cite{Fischer2012}. This alloy composition and density are consistent with seismological measurements, which suggest that Earth's core is approximately 10\% under-dense compared to pure iron \citep{Birch1952, Stevenson1981, Souriau2007}. Moreover, the requirement for a long-lived geodynamo further suggests a liquid, convecting outer core consistent with liquid iron alloys \citep{Stevenson1981, Jeanloz1990, McDonough1995}. Figure~\ref{fig:fig2} shows the density and heat capacity differences of the Fe$_{16}$Si EOS of \cite{Fischer2012} and the liquid iron EOSes of \cite{Ichikawa2014} and \cite{Dorogokupets2017Thermodynamics6000K}. The Fe$_{16}$Si EOS is $\sim$10\% less dense than pure iron, and its heat capacity is nearly half that of pure iron. 

\subsection{Conductivities \& Opacities} \label{subsec:cond_opa}

\begin{figure}[ht!]
\centering
\includegraphics[width=0.43\textwidth]{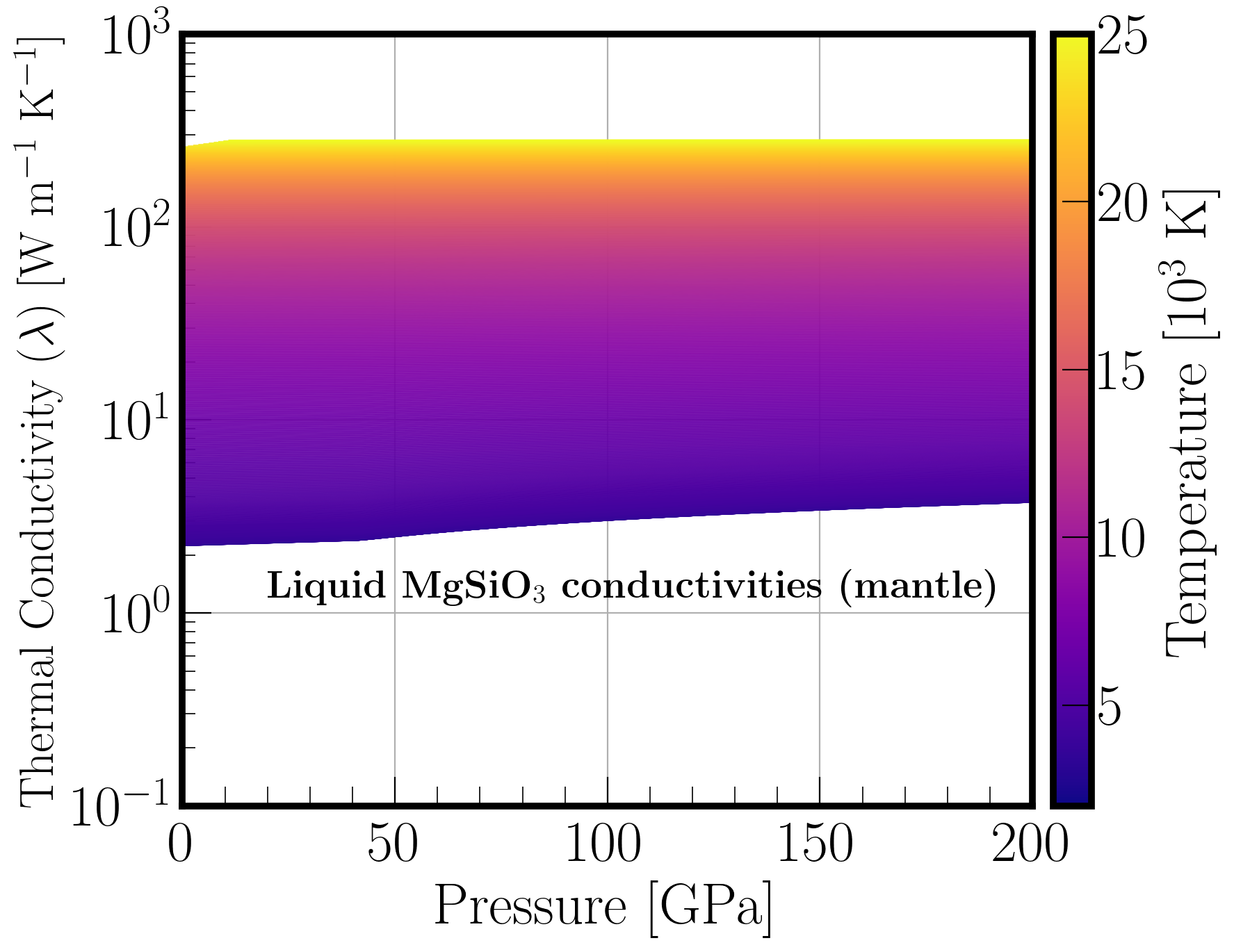}
\includegraphics[width=0.43\textwidth]{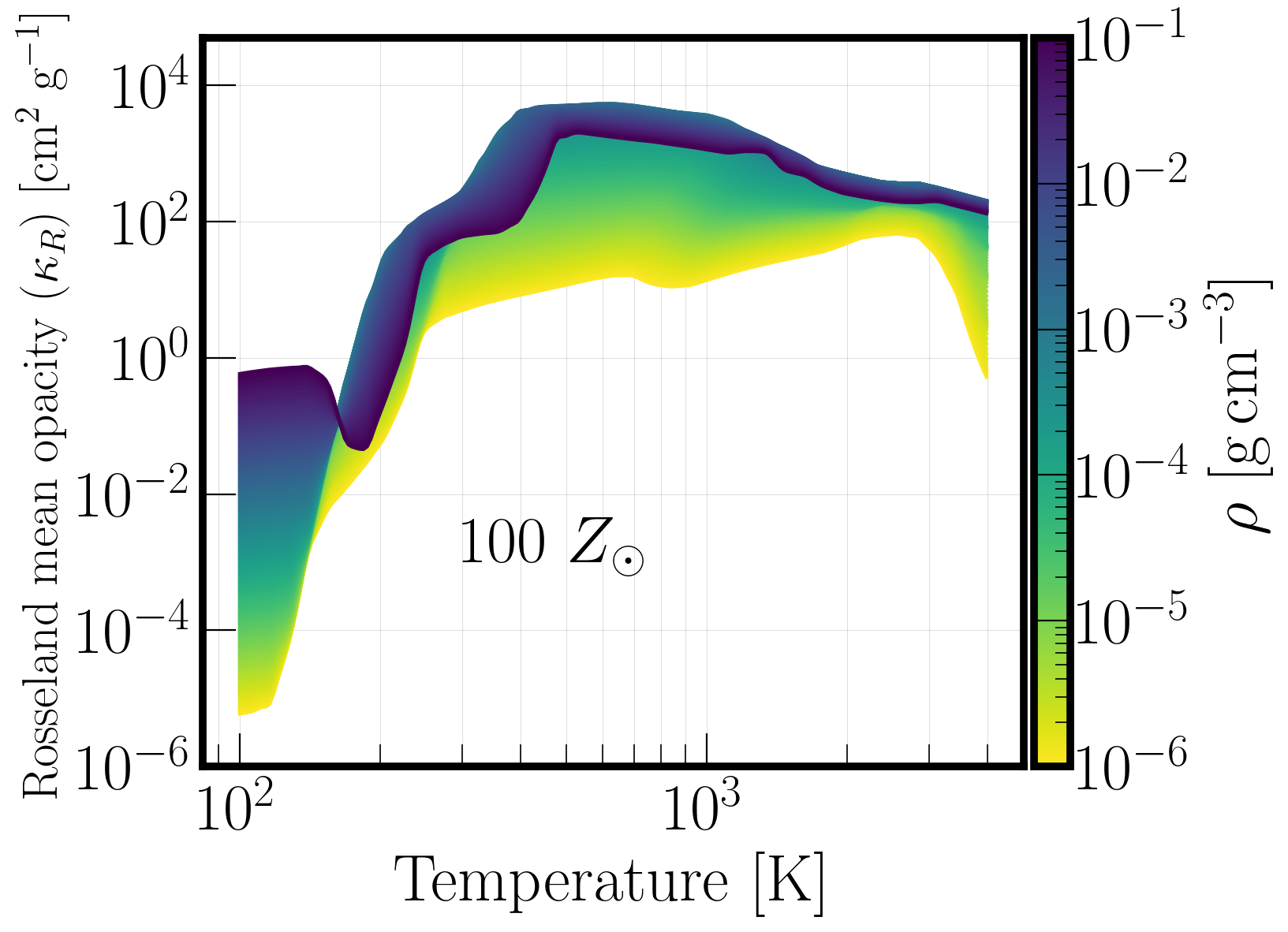}
\caption{The thermal conductivity of MgSiO$_3$(top) implemented in the mantle, which governs non-convective heat transport. The electronic and phonon (lattice or vibrational) components are included \citep{Stamenkovic2011, PengDeng2024}. The electronic component dominates at higher temperatures. The Rosseland mean opacities (bottom) at 100 times solar abundance control the radiative heat transport in the envelope \citep{Sharp2007, LacyBurrows2023}. All of these quantities are density- and temperature-dependent. The thermal conductivities reported in \cite{French2019} govern heat conduction in the envelope when water is used. }
\label{fig:fig3}
\end{figure}

We use the updated density- and temperature-dependent thermal conductivities of MgSiO$_3$ as given by \cite{PengDeng2024} for conductive heat transport in the mantle. The electronic contribution to the thermal conductivity is given by \cite{Stamenkovic2011}.\footnote{The electronic and radiative parts of the thermal conductivity of MgSiO$_3$ remain unconstrained. The conductivities of \cite{Stamenkovic2011} should be regarded as estimates.} Shown in the top panel of Figure~\ref{fig:fig3}, the thermal conductivity depends more strongly on temperature than pressure. In the core, we use a constant thermal conductivity of 40 W m$^{-1}$ K$^{-1}$, which is consistent with the thermal conductivity of Earth's core \citep{Pozzo2022, Luo2024}. The thermal conductivities of the envelope are those of \cite{French2019}, combined with the H-He thermal conductivities used in gas giants \citep[See Figure 1 of][]{Sur2024a} when the envelope metals involve water mixtures. 

We use updated Rosseland mean opacities ($\kappa_R$) ranging from 3 to 100 times the solar metallicity \citep{Sharp2007, LacyBurrows2023} in the envelope. These span a range of 100--4000 K in temperature and a density range of 10$^{-5}$ to 10$^{-2}$ g cm$^{-2}$. We interpolate between them to obtain $\kappa_R(\rho, T, Z)$ in the envelope, where $\rho, T, Z$ are the local mass densities, temperatures, and metal mass fractions. When the metal composition of the envelope changes due to convective mixing of stably-stratified regions or miscibility (See Sections~\ref{subsec:stratification}, \ref{subsec:miscibility}) the opacity metallicity adapts to the interior metal content, therefore changes with time. Example conductivities and opacities are shown in Figure~\ref{fig:fig3}, which illustrate the relevant ranges of transport properties for the mantle and the envelope. 

\subsection{Atmosphere Model} \label{subsec:atm}
We use a non-gray radiative-convective equilibrium atmospheric model \citep{Fortney2005, Fortney2007, Fortney2020, Ohno2023} with updated grids calculated by and implemented by \cite{Chachan2025a} and \cite{Tang2025}. These atmosphere models cover a range of $\log{g/\rm g\ cm^{-2}} \in$ 1--5, a metallicity range from 1 to 100 times the solar metallicity, and a stellar flux coverage from 0.73 to 1000 F$_\oplus$, where F$_\oplus$ is the present incident flux from the Sun to the Earth of 1361 W m$^{-2}$. For this work, we do not exceed stellar fluxes higher than 1000 F$_\oplus$. These atmosphere model tables make the \textit{intrinsic} temperature ($T_{\rm int}$), which is the temperature associated with the interior flux, dependent on the temperature at a pressure of 1kbar ($T_{1\rm kbar}$)\footnote{For example, others use the temperature at 1 or 10 bars, such as those from \cite{Fortney2011}.}, the metallicity, and instellation (or stellar incident flux). The instellation is calculated from the equilibrium temperature ($T_{\rm eq}$), a free parameter in our models. Essentially, $T_{\rm int} = T_{\rm int}(\log g, T_{1\rm kbar}, T_{\rm eq}, Z_{1\rm kbar}, \mathcal{F_*}$)\footnote{The metal mass fraction at 1kbar ($Z_{1\rm kbar}$) is converted to a solar ratio metallicity before passing it to the atmosphere model.}, where $\mathcal{F}_*$ is incident stellar flux. The metallicity is converged from the local metal mass fraction. The effective temperature of our models is obtained using the relation $T_{\rm eff}^4 = T_{\rm eq}^4 + T_{\rm int}^4$.

Changes in the intrinsic temperature control the envelopes' cooling. As such, we predict the cooling rate of the next timestep by including the partial derivatives of the intrinsic temperature via the chain rule:

\begin{equation}
\frac{\partial T_{\rm int}}{\partial S_{1\rm kbar}} = 
\underbrace{\bigg(\frac{\partial T_{\rm int}}{\partial T_{1\rm kbar}}\bigg)}_{\text{from atmosphere model}} 
\underbrace{\bigg(\frac{\partial T_{1\rm kbar}}{\partial S_{1\rm kbar}} \bigg)_P}_{\text{from EOS}},
\end{equation}
where $S_{1\rm kbar}$ is the local entropy of the profile at 1kbar pressure. The temperature at 1kbar pressure of EOS partial derivative is a function of entropy and pressure, $T(S, P)$, inverted from $S(P, T)$ \citep{Tejada2024}, and it is taken at a constant pressure.  The adiabatic gradients of this atmosphere model are calculated using the H-He EOS of \cite{Chabrier2019} \citep{Fortney2020}, which is identical to the H-He EOS used here \citep{Chabrier2021} at low pressures, so the EOS derivatives are self-consistent. This derivative is included in Eq.~C6 of \cite{Sur2024a} to treat the energy flux of the atmosphere boundary condition implicitly.

Throughout most of this work, the ``radius'' is the radius at 1 bar (R$_{\rm 1 bar}$). The pressure of 1 bar is our upper boundary condition for hydrostatic equilibrium \citep[Eq. 6 in][]{Sur2024a}. More relevant to observations is the transit radius (R$_{\rm transit}$), which is used in Sections~\ref{subsec:miscibility} and \ref{subsec:case_studies}. The transit radius is estimated using the isothermal, two-stream radiative transfer model of \cite{Guillot2010},

\begin{equation} \label{eq:rtransit}
    R_{\rm transit} = R_{\rm 1 bar} + \Delta z
\end{equation}
where 
\begin{equation} \label{eq:deltaz_guillot}
    \Delta z = H(r) \ln\bigg[\gamma\bigg(\frac{2\pi r}{H(r)}\bigg)\bigg].
\end{equation}

Here, $H(r) = k_B T_{\rm irr}/\mu m_pg$ is a characteristic scale height where $T_{\rm irr}$ is the irradiation or equilibrium temperature, $\mu$ is the mean molecular weight, $g$ is the gravitational acceleration at 1 bar, and $\gamma = 0.6\sqrt{T_{\rm irr}/2000 K}$ \citep{Guillot2010, Rogers2011}. Typical $\Delta z$/R$_{\rm 1 bar}$ are 1-2\% in our models, but this could be beyond 30\% for low-gravity, highly-irradiated sub-Neptunes \citep[see e.g.,][]{HoweBurrows2015, Tang2025}.

\subsection{Heat and Compositional Transport} \label{subsec:transport}

Heat transport in \texttt{APPLE} is modeled by a combination of radiative, conductive, and convective heat transport methods. The total flux is $\mathcal{F}_{\rm tot} = \mathcal{F}_{\rm rad} + \mathcal{F}_{\rm cond} + \mathcal{F}_{\rm conv}$, where $\mathcal{F}_{\rm rad}$ is the diffusive radiative heat flux, $\mathcal{F}_{\rm cond}$ is the conductive heat flux, and $\mathcal{F}_{\rm conv}$ is the convective heat flux. 

The radiative flux is modeled with radiative diffusion, and is relevant for the radiative cooling of the envelope, and is given by \citep[e.g.,][]{Kippenhahn1990, Sur2024a}

\begin{equation}\label{eq:rad}
    \mathcal{F}_{\rm rad} = -\frac{4ac}{3}\frac{T^3}{\kappa_R}\frac{\partial T}{\partial r},
\end{equation}
where $\partial T/\partial r$ is the temperature gradient. 

The conductive heat flux is given by 

\begin{equation}\label{eq:cond}
    \mathcal{F}_{\rm cond} = -\lambda\frac{\partial T}{\partial r},
\end{equation}
where $\lambda$ is the thermal conductivity (See top panel of Figure~\ref{fig:fig3}).  

We updated \texttt{APPLE} to account for the viscous-limited MLT formalism, described in this section. A modified MLT prescription is applied for regions under the melting line of MgSiO$_3$ \citep[][ZR22 hereafter]{Zhang2022}.
 
The convective flux, $\mathcal{F_{\rm conv}}$, may be written as 

\begin{equation}\label{eq:fconv}
    \mathcal{F_{\rm conv}} = -\rho T\kappa_h \frac{\partial S}{\partial r},
\end{equation}
which is equivalent to the Schwarschild-limited convective flux described in Equations 22 and 25 of \cite{Sur2024a} used in \texttt{APPLE} for gas giant planet evolution. In Eq.~\ref{eq:fconv}, $\kappa_h$ is the eddy diffusivity, as described in ZR22, and $\partial S/\partial r$ is the specific entropy gradient \citep[See Section 6 and Appendix A of][for a derivation from the traditional temperature gradient formalism.]{Tejada2024}. 

In the traditional inviscid regime, the eddy diffusivity is the convective flux coefficient, given by

\begin{equation}\label{eq:inv_eddy}
    \kappa_h = \sqrt{\frac{\alpha g T l^4}{32 C_p}\frac{\partial S}{\partial r}}
\end{equation}
where $\alpha$ is the thermal expansion coefficient, $l$ is the mixing-length parameter, and $C_p$ is the isobaric specific heat capacity. We assume no compositional change in the mantle and thus apply a Schwarzschild convection criterion for convection ($\partial S/\partial r < 0$). This is expanded to the Ledoux criterion \citep{Ledoux1947} in the presence of compositional gradients in the envelope,  

\begin{equation}\label{eq:ledoux}
    \frac{dS}{dr} - \sum_i\frac{\partial S}{\partial X_i}_{\rho, P}\frac{dX_i}{dr} < 0,
\end{equation}
applied in Sections~\ref{subsec:stratification} and \ref{subsec:miscibility}. In Eq.~\ref{eq:ledoux}, $X_i$ are helium and heavy element mass fractions. The entropy formalism of the Ledoux criterion is derived in \cite{Lattimer1981} and \cite{Tejada2024}.

In cases where viscous drag forces are significant, the fluid velocity is constrained by the viscosity of the material in the mantle \citep[See Appendix A of][for a physical derivation]{Zhang2022}. Thus, when the mantles cool sufficiently to the solid phase, we apply a modified MLT prescription \citep{Sasaki1986}.  In this limit of viscous convection, the eddy diffusivity is expressed as

\begin{equation}\label{eq:visc_eddy}
    \kappa_h = \frac{\alpha g T l^4}{32 C_p \nu }\frac{\partial S}{\partial r},
\end{equation}
where $\nu$ denotes the kinematic viscosity, obtained from the dynamic viscosity via $\nu = \eta / \rho$. To calculate the dynamic viscosity, $\eta$, we adopt an Arrhenius formulation for the kinematic viscosity, as found in Eq. 41 of ZR22 as obtained by \cite{Ranalli2001}. Following ZR22, we assume that the viscosity of liquid MgSiO$_3$ is 100 Pa s \citep{Abe1997}, which is a typical value for basaltic lava on Earth \citep{Harris2008}.\footnote{On Earth, rhyolitic (silicate-rich) lava has a viscosity ranging from 0.1 to 10$^{14}$ Pa s, depending on its silicate fraction \citep{Giordano2008_magma}. Moreover, \cite{Luo2025} calculated viscosities for the EOS used here, finding much lower values of roughly $\sim$10$^{-2}$ Pa s (see their Figure 9). We choose 100 Pa s here due to its precedence in rocky planet evolution models and for convenience. } In practice, this viscosity is only used in the transition regions of the mantle between liquid and solid states, since the inviscid limit is not dependent on the viscosity (Eq.~\ref{eq:inv_eddy}). 

In traditional MLT, then,

\begin{equation}
    \mathcal{F_{\rm conv}} \propto (\partial S/\partial r)^{3/2}
\end{equation}
since Eq.~\ref{eq:inv_eddy} contains a squared-root spatial derivative. In modified MLT \citep[e.g.,][]{Sasaki1986} for viscous convection, however, 

\begin{equation}
    \mathcal{F_{\rm conv}} \propto (\partial S/\partial r)^{2}
\end{equation}
given Eq.~\ref{eq:visc_eddy}.

The transition between the inviscid and viscous convection limits (Eqs. \ref{eq:inv_eddy} and \ref{eq:visc_eddy}) is determined by the local melt fraction of the silicate mantle, $\chi$. We define the melt fraction in terms of the melting temperature, $T_m$ obtained from \cite{Fei2021}, and a temperature transition width ($\Delta T_{\rm lat}$), which we set at 150 K for numerical time-stepping stability during the liquid-solid transition,

\begin{equation}\label{eq:chi}
  \chi(P,T) = \tfrac{1}{2}\!\left[1+\tanh\!\left(\frac{T - T_{\mathrm{m}}(P)}{\Delta T_{\mathrm{lat}}}\right)\right].
\end{equation}

Thus, $\chi$ approaches 1 for pure liquid states and 0 for pure solid states, ensuring that $\chi$ is differentiable with the local pressure and temperature. We interpolate the viscosity log-linearly, weighted by the melt fraction, as done by ZR22. For partially melted regions, where $0 <\chi< 1$, we incorporate Eqs. 34--38 from ZR22 to smoothly transition from the inviscid convection limit (Eq. \ref{eq:inv_eddy}) to the viscous convection limit (Eq. \ref{eq:visc_eddy}). 

We treat latent heating as an additional source term ($L_{\rm lat}$) added to the luminosity term ($L$) in the discrete entropy equation for each shell $k$. Starting from

\begin{equation}\label{eq:energy}
\rho T \,\frac{dS}{dt}
=
-\,\frac{\partial L}{\partial m}
+
\rho\,L_{\mathrm{lat}}\,\frac{d\chi}{dt} + ...,
\end{equation}
we discretize in time from step $n$ to $n+1$ in the energy equation as \citep[see Appendix C of ][]{Sur2024a}

\begin{equation}
    \begin{split}
         \bigl(S_k^{n+1} - S_k^{n}\bigr)
        +
        \frac{\Delta t}{m_k\,T_k^{n+1}}
        \left(L_{k+\frac12}^{n+1} - L_{k-\frac12}^{n+1}\right)
        + \\
        \frac{L_{\mathrm{lat}}}{T_k^{n+1}}\,
        \bigl(\chi_k^{n+1} - \chi_k^{n}\bigr) + ...
        = 0,
    \end{split}
\end{equation}
where $m_k$ is the mass in shell $k$, $T_k^{n+1}$ is the updated temperature, and $L_{k\pm 1/2}^{n+1}$ are the total (convective + radiative) luminosities at the shell boundaries. The local melt fraction $\chi(P, T)$ is evaluated implicitly at the new time level using a smooth transition across the melt curve (Eq.~\ref{eq:chi}) so that the latent term depends on $T_k^{n+1}$ and enters the global Newton-Raphson solve. In the Jacobian, we include the derivative $d\chi/dT$, which contributes a local factor $L_{\mathrm{lat}}(d\chi/dT)/C_p$ to the entropy equation. This behaves like an enhanced effective heat capacity in partially molten cells, ensuring that energy spent on melting or freezing is self-consistently accounted for and that zones crossing the melt curve cool more slowly.

We assume that the latent heat, $L_{\rm lat}$, release value in the silicate mantle is $7.322 \times 10^5$ J kg$^{-1}$ \citep{Hess1990KomatitiesMantle}. The latent heat release in the iron core is $1.2 \times 10^6$ J kg$^{-1}$ \citep{Anderson1997}. To release the latent heat of the iron-rich core, we use the iron phase transition of \cite{Zhang2015}, as done in ZR22, but we note here that iron alloy melting curves could be much lower \citep{Ezenwa2024}.

The internal radiogenic heat released in the mantle and core is given by 

\begin{equation}
H(t)
  = \sum_{i}
    w_i\,q_{0,i}\,\exp\!\left[\ln 2\,\bigg(-\frac{t}{\tau_i}\bigg)\right],
\label{eq:Hspec}
\end{equation}
where $i$ represents the radioactive species of $^{40}\rm K,\,^{232}\rm Th,\,^{235}\rm U,\,^{238}\rm U$, $q_{0,i}$\footnote{See Table 4 of ZR22 for values used here.} are the present-day heat production rates corresponding to each species per unit Earth's mantle mass, $\tau_i$ the decay times of such species, and $t$ is the current model age. We assume that the abundances of each radioactive species are those of Earth's mantle \citep{McDonough1995}, scaled to sub-Neptune mantle masses, and are evenly distributed uniformly throughout the interior, as done for super-Earth interiors in ZR22.

Convection-dominated regions mix material and homogenize the composition in timescales shorter than evolutionary timescales. Thus, convective regions will exhibit flat composition profiles wherever the diffusion equation, 

\begin{equation}\label{eq:conv_mixing}
    \frac{\partial X_i}{\partial t} = \frac{\partial}{\partial M_r}\bigg(4\pi r^2 \rho \mathcal{D}\frac{\partial X_i}{\partial r}\bigg),
\end{equation}
is applied. Here, $M_r$ is the mass shell at radius $r$, $\mathcal{D} = \frac{1}{3}v_{\rm MLT}l$ is the convective diffusion coefficient defined by the convective velocity, $v_{\rm MLT}$, the local mixing length, $l$, and $\rho$ is the mass density \citep[Eqs. 44 and 45 in][]{Sur2024a}. This MLT diffusion coefficient is added to the self-diffusion coefficients of H-He-$Z$ mixtures (typically $\sim$$10^{-3}$--$10^{-4}$ cm$^2$ s$^{-1}$). 

\begin{figure*}[ht!]
\centering
\includegraphics[width=0.32\textwidth]{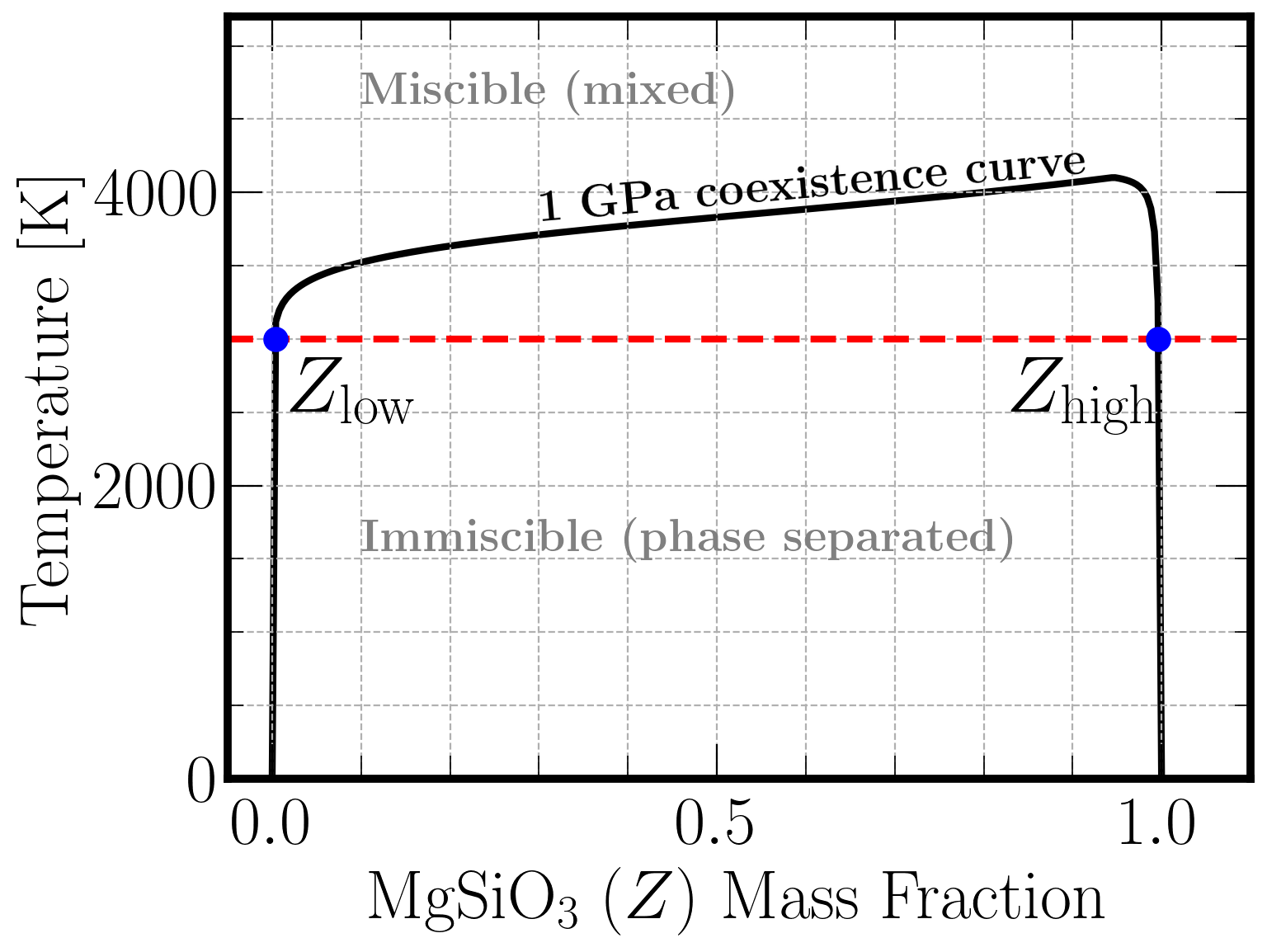}
\includegraphics[width=0.32\textwidth]{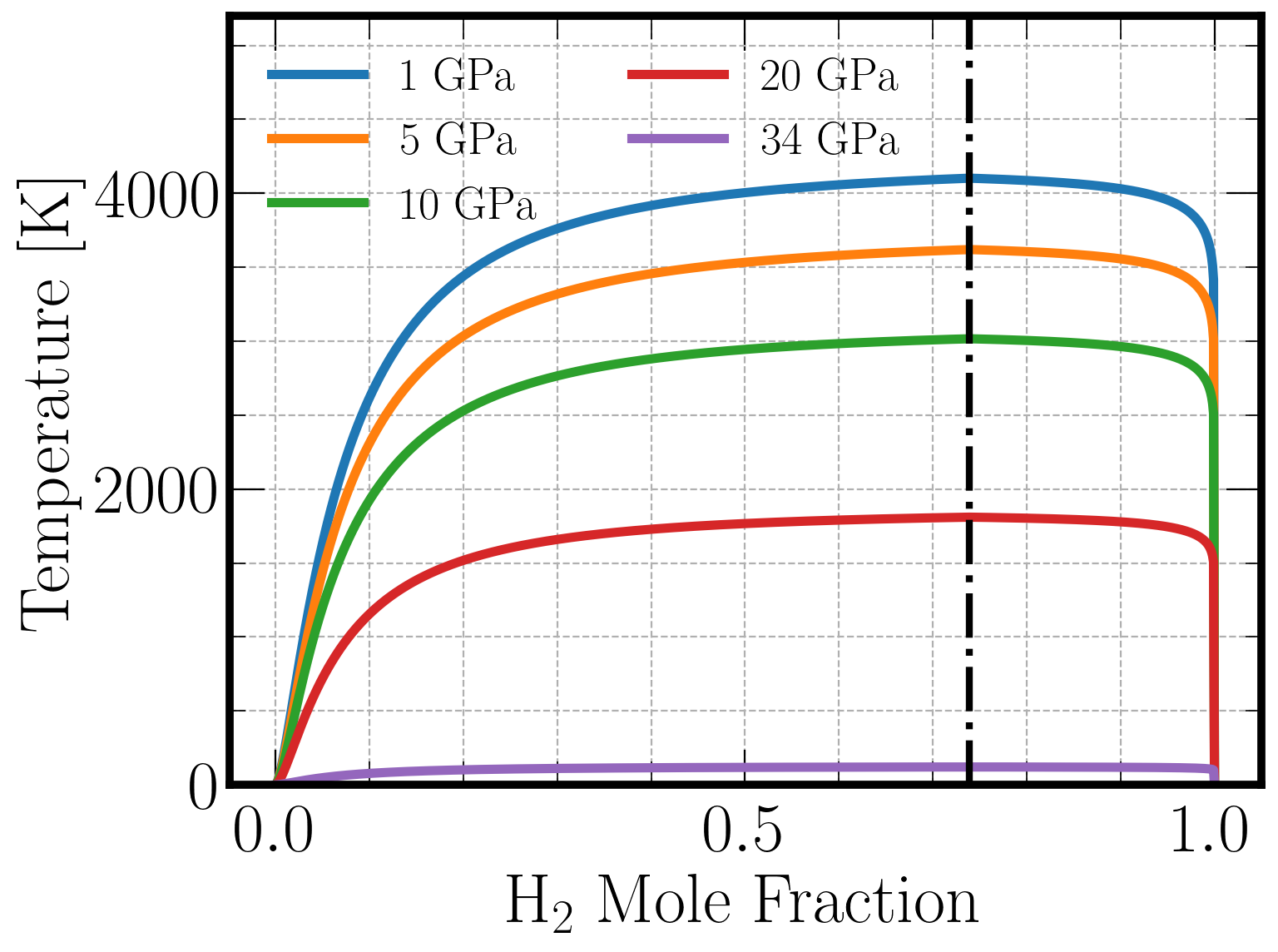}
\includegraphics[width=0.32\textwidth]{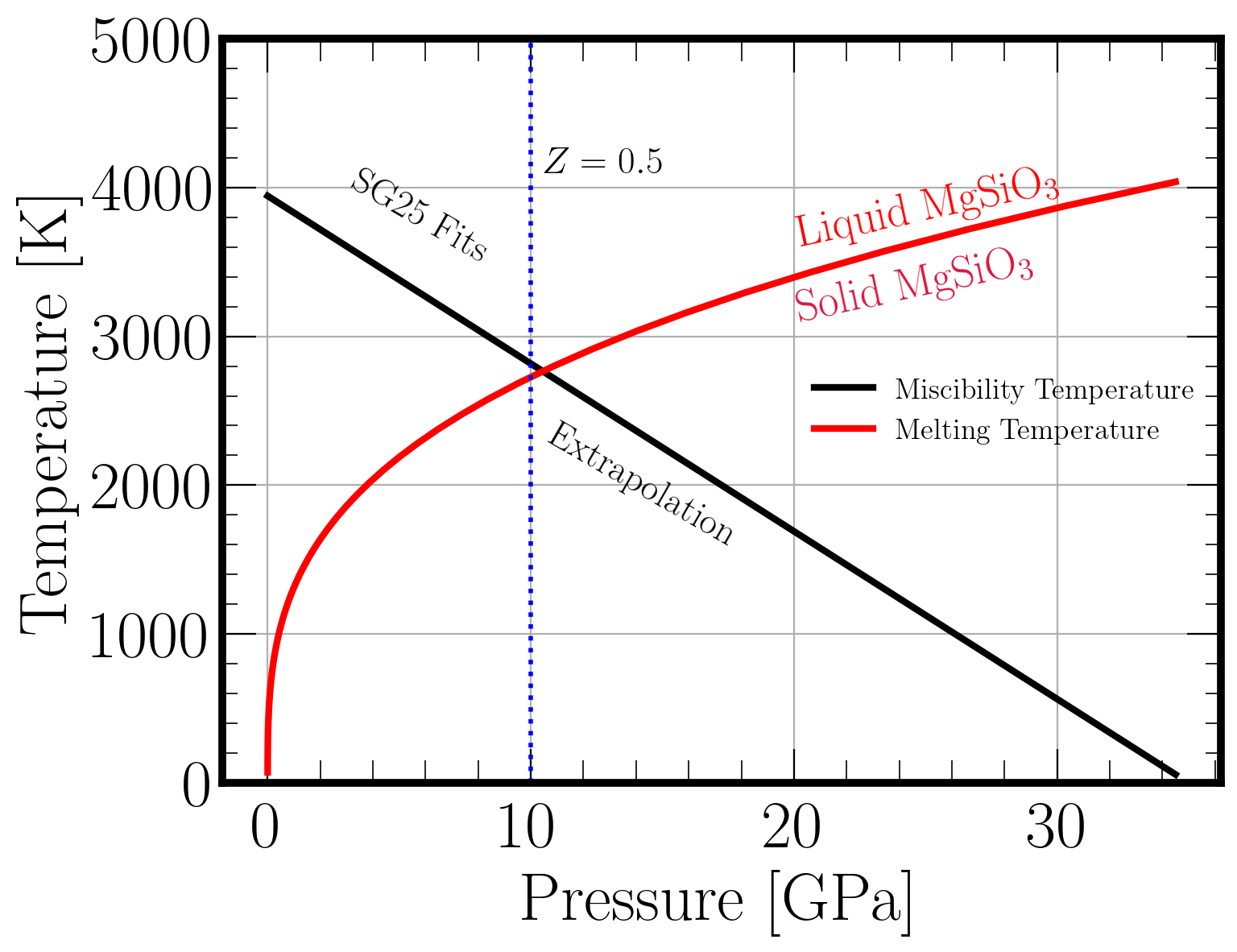}
\caption{Coexistence (binodal) curve temperature curve fits of \citet[][SG25]{StixrudeGilmore2025} are shown in the left and center panel as a function of silicate mass fraction and H$_2$ mole fraction, respectively. In the left panel, a given local temperature (e.g., the red dashed line) can intersect the coexistence curve at two equilibrium abundances, these being $Z_{\rm low}$ and $Z_{\rm high}$. The diffusion-advection method described by Eq.~\ref{eq:misc} uses $Z_{\rm low}$ and $Z_{\rm high}$ to drive the local metal fraction to $Z_{\rm low}$ and $Z_{\rm high}$. The center panel shows coexistence curves at constant pressure from 1 to 34 GPa. Above the critical temperatures (the maximum temperatures of each coexistence curve), mixtures remain miscible regardless of composition. At temperatures below the critical temperature, mixtures remain miscible outside the coexistence curves. The dash-dot line traces the critical points, i.e., (critical composition, critical temperature) of the respective coexistence curves. The right panel shows the miscibility temperatures at a constant silicate fraction of 0.5 compared with the melting curve of \cite{Fei2021}, shown in red. Higher pressures lead to lower miscibility temperatures, eventually crossing the melting curve of MgSiO$_3$ (shown in red) at 10 GPa (depicted by the vertical dotted line), as shown in the right panel. As such, extrapolations beyond this point are likely unphysical.}
\label{fig:fig_coex_curves}
\end{figure*}

For silicate-hydrogen miscibility composition transport, relevant to the models presented in Section~\ref {subsec:miscibility}, we adopt the diffusion-advection scheme B described by \cite{Sur2024a}, originally designed for helium rain in gas giant planets. We generalize Eq. 49 of \cite{Sur2024a} incorporating it to $Z$ advection-diffusion,

\begin{equation}\label{eq:misc}
    \begin{aligned}
        \frac{\partial Z}{\partial t} 
        &= \frac{\partial}{\partial M_r}\Bigg[
        4\pi r^2 \rho \mathcal{D}\Bigg(
        \frac{\partial Z}{\partial r}
        + \frac{\max(0, Z-Z_{\rm low})}{\mathcal{H}_r}
           \frac{\partial Z_{\rm low}}{\partial T}\Big|_P \\
        &\qquad\quad
        + \frac{\max(0, Z-Z_{\rm high})}{\mathcal{H}_r}
           \frac{\partial Z_{\rm high}}{\partial T}\Big|_P
        \Bigg)\frac{T}{C_P}\Delta S\Bigg]
    \end{aligned}
\end{equation}
where $\mathcal{H}_r$ is the rain scale height (typically 100 km), and $\Delta S$ is the change in specific entropy due to cooling. The metal mass fractions $Z_{\rm low}$ and $Z_{\rm high}$ correspond to the equilibrium compositions on the two branches of the coexistence (or binodal) curve intersected by a given pressure–temperature profile at the local bulk composition, as illustrated in Figure~\ref{fig:fig_coex_curves}. The coexistence surface delineates regions in the pressure–temperature–composition space in which a binary mixture is either fully miscible or undergoes phase separation into two compositionally distinct phases, with the latter defining the miscibility gap. A slice through this surface at fixed pressure (or temperature) yields the coexistence (binodal) curve at that pressure (or temperature); see the middle panel of Figure~\ref{fig:fig_coex_curves}. By contrast, the critical curve is the locus of critical points marking the lowest temperatures at each pressure above which the binary mixture remains homogeneous for all compositions. Finally, a miscibility curve is the locus in pressure–temperature space at a fixed composition that separates single-phase stability from two-phase coexistence.

The left panel of Figure~\ref{fig:fig_coex_curves} shows the coexistence curve temperatures of \cite{StixrudeGilmore2025} at 1 GPa for demonstration purposes as a function of MgSiO$_3$ mass fraction and pressure. A given temperature will intercept the coexistence curve at two equilibrium points, $Z_{\rm low}$ and $Z_{\rm high}$, which are located at each side of the critical temperature. This is illustrated by the intercept marks on either side of the coexistence curve at 3,000 K (dashed red line). We invert this miscibility curve to obtain $Z_{\rm low}(P, T)$ and $Z_{\rm low}(P, T)$, where $P,T$ are any pressure-temperature coordinates along the envelope. If no solution is found for either of the equilibrium abundances, indicating fully miscible regions, these are set to 1.0, eliminating the advection terms in Eq.~\ref{eq:misc}. The center panel shows various coexistence curves as a function of hydrogen mole fraction. The right panel of Figure~\ref{fig:fig_coex_curves} shows that pressures above 10 GPa are extrapolations from the calculations and fits of \cite{StixrudeGilmore2025}, since these go below the melting temperatures of MgSiO$_3$ \citep{Fei2021}. As shown in the right panel of Figure~\ref{fig:fig_coex_curves}, the silicate-hydrogen miscibility curve fits (in black) have a negative temperature-pressure gradient, implying that profiles only intercept in one region of the ``binodal surface'' \citep[see left panel of Figure 4 of][]{Rogers2025}. This means that the miscibility region is unbounded above the upper envelope so that silicate rain would occur everywhere above this single intercept. In practice, for numerical stability, we impose a lower limit of 0.1 GPa to define the lower bound of the silicate rain region, extending up to 35 GPa. 

We highlight here that the local metal abundance, $Z$, is driven to either $Z_{\rm low}$ or $Z_{\rm high}$ depending on their location in the interior structure, and the rate at which we rain the local $Z$ is proportional to the difference between $Z$ and its equilibrium value (e.g., $Z-Z_{\rm low}(P, T)$). This means that phase separation is not instantaneous, as it depends on the distance from the local coexistence curves and the rain scale height, $\mathcal{H}_r$. The partial derivatives of the equilibrium abundances with respect to local temperature, evaluated at constant local pressure, are used to inform our implicit Jacobian update.  

\section{Evolution of Sub-Neptunes}\label{sec:evolution}

In the context of giant planets, assuming that their envelopes are fully adiabatic and homogeneous, eliminates all traces of initial conditions at evolutionary timescales ($\gtrsim$500 Myr) due to efficient adiabatic cooling. This means that adiabatic models have no memory of their initial conditions at late times. However, the treatment of initial conditions becomes essential when modeling the interior evolution of non-adiabatic and inhomogeneous interiors, such as stably-stratified regions in the Solar System gas and ice giant planets \citep{Nettelmann2013a, Nettelmann2016, Vazan2018, Scheibe2021, Knierim2024, Tejada2025, Sur2025a, Tejada2025b}. This implies that inhomogeneous and non-adiabatic interiors may retain some memory of their initial conditions, since the final states are sensitive to them \citep{Tejada2025, Sur2025a, Knierim2025b}. The following sections demonstrate these effects in the interior of sub-Neptunes.

\subsection{Mantle Initial Temperature Dependencies}\label{subsec:entropy}

We demonstrate in this section that hotter initial sub-Neptune mantles and cores cool slowly due to limited heat flux transfer across the envelope-mantle boundary (EMB), leading to a 5--10\% increase in the radii at late ages compared to models with cooler starts. This is due to the steep mean molecular-weight gradient between the mantle and the envelope, which is stable against Ledoux convection. Heat, therefore, can only be transported via conduction and radiation. In the standard MLT approach to convection \citep{Bohmvitense1958MLT}, heat is transported efficiently due to fluid parcels being dispersed over a characteristic length, whereupon the parcels deposit their heat and equilibrate with their surroundings. At the EMB, $\mathcal{F}_{\rm tot} = \mathcal{F}_{\rm cond}$, since fluid parcels cannot cross the EMB. Eq.~\ref{eq:cond} therefore implies that

\begin{equation}
    \frac{\partial T}{\partial r} \sim \frac{L}{4 \pi r^2\lambda},
\end{equation}
indicating that the temperature gradient must be large since the conductivities are small ($\sim$5--7 W m$^{-1}$ K$^{-1}$) at the EMB. This, in turn, causes shallow temperature gradients within the mantle, as heat is only conducted. Shallow temperature gradients in the mantle make $\partial S/\partial t$ (Eq.~\ref{eq:energy}) small and, hence, cooling is slowed. 

\begin{figure*}[ht!]
\centering
\includegraphics[width=0.47\textwidth]{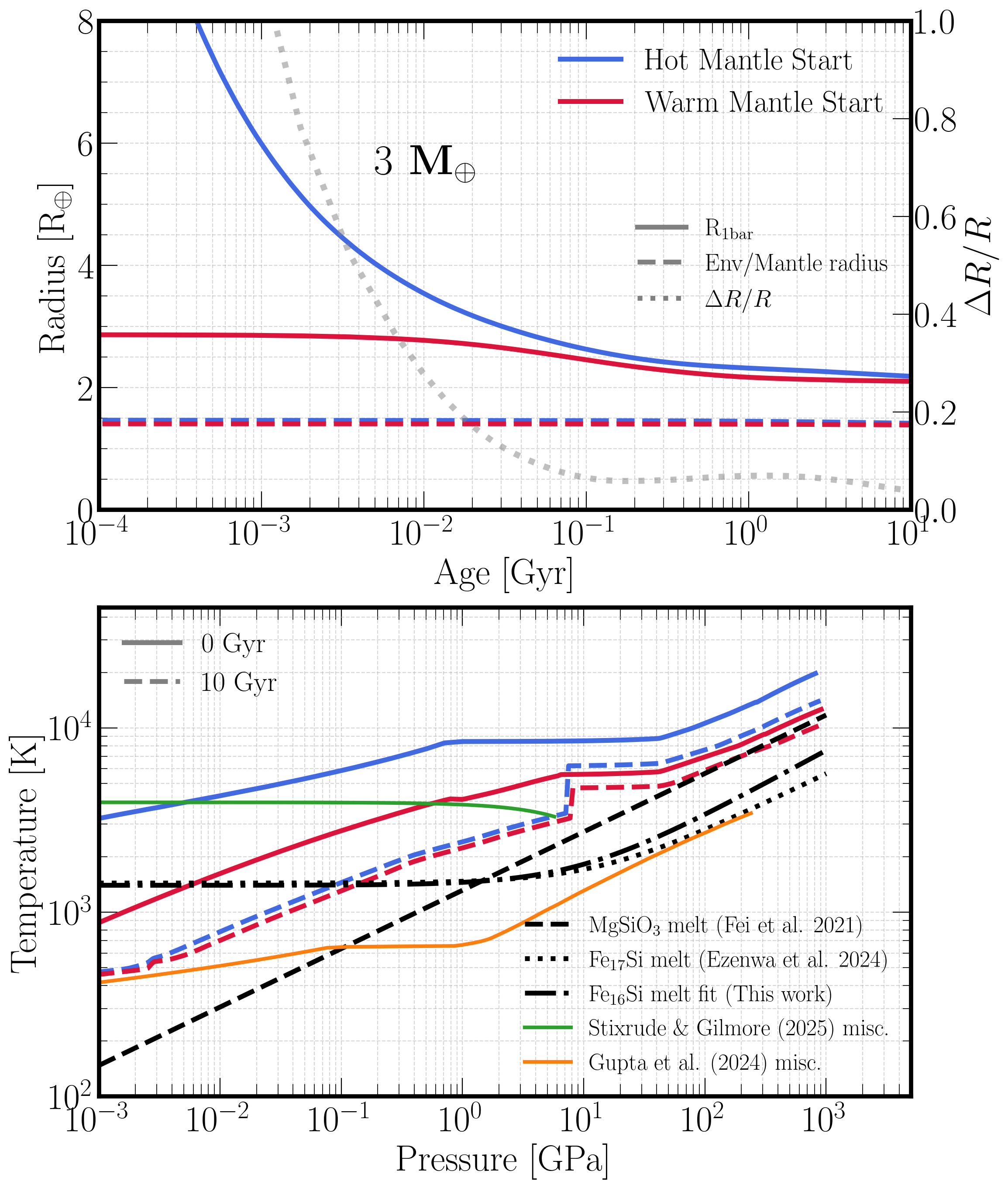}
\includegraphics[width=0.47\textwidth]{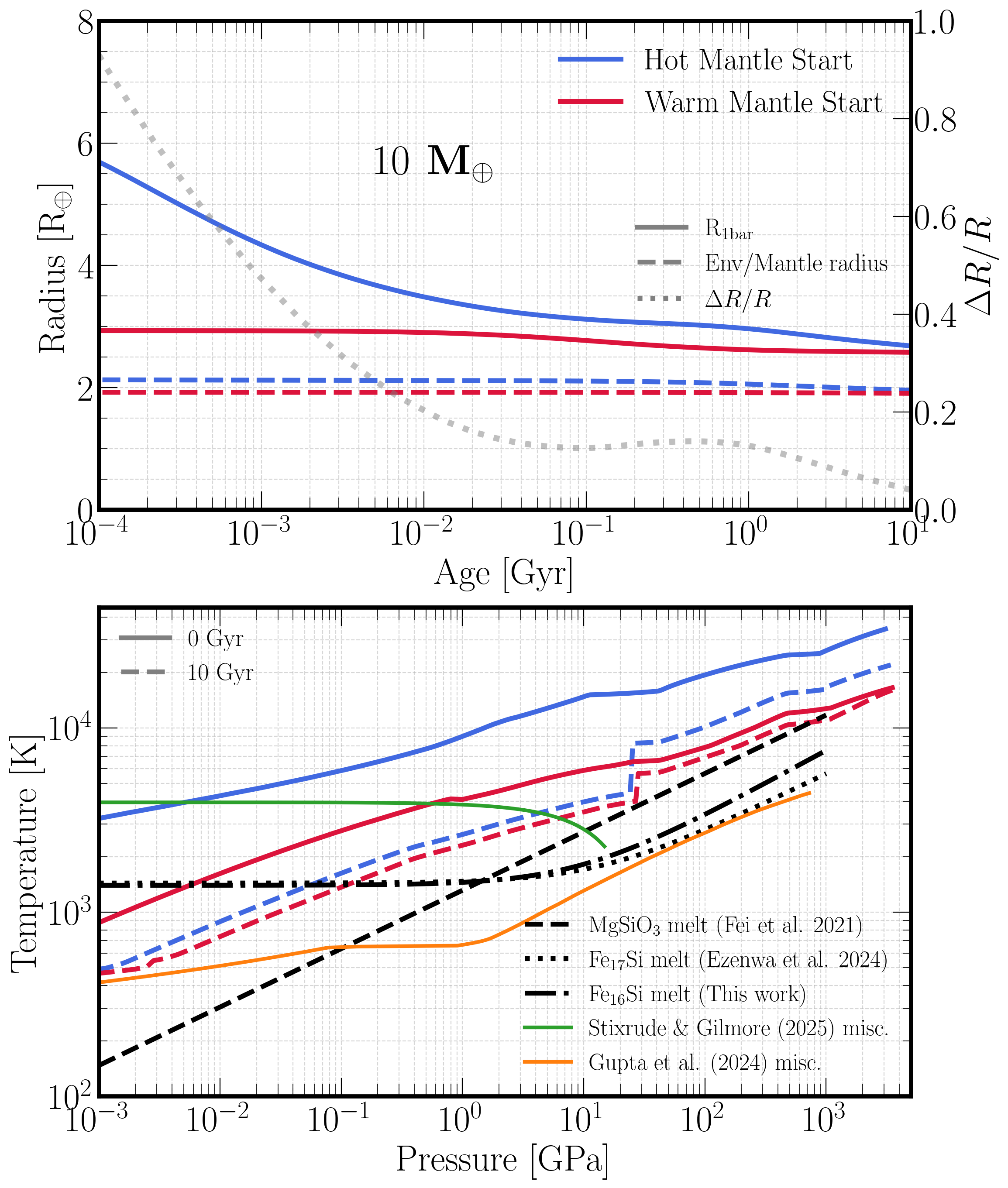}
\caption{Impact of initial mantle thermal state on the evolution of 3 and 10 M$_\oplus$ sub-Neptunes. Hotter initial mantles retain heat, maintaining significantly inflated radii even at late ages. These effects are more significant for more massive sub-Neptunes. Models assume an equilibrium temperature of 400 K, an envelope mass fraction of 5\%, envelope metal mass fraction of $0.5$ (water), and a 1:2 core-to-mantle ratio. The top panel shows the radius evolution of models initialized with hot mantle temperatures of 9,000 K and 16,000 K for the 3 and 10 M$_\oplus$ models, respectively, compared with models initialized at mantle temperatures of 5,000 K and 7,000 K, respectively. Solid lines in the top row show the 1-bar radius; dashed colored lines mark the envelope–mantle boundary. The gray dotted line tracks the fractional radius difference ($\Delta R/R$). The bottom row shows pressure-temperature profiles at 0 Gyr (solid) and 10 Gyr (dashed). Over-plotted are miscibility curves for hydrogen-silicates (green; \citealt{StixrudeGilmore2025}) and hydrogen-water (orange; \citealt{Gupta2025}), alongside melting curves for MgSiO$3$ \citealt{Fei2021}. The melting curves of Fe-alloys Fe$_{17}$Si \citep[black dotted;][]{Ezenwa2024} and Fe$_{16}$Si \citep[black dash dot;][]{Fischer2012} are also plotted; see Section \ref{subsec:entropy} for details.}
\label{fig:fig_3_10_evol}
\end{figure*}

We demonstrate this effect in Figure \ref{fig:fig_3_10_evol}, where we show the evolution of 3 and 10 M$_\oplus$ sub-Neptune models starting at hot ($\gtrsim$10,000 K) and cold ($\sim$7,500 K) temperatures at the EMB, shown in solid color lines.\footnote{The specific temperatures here are of less importance. What Figure~\ref{fig:fig_3_10_evol} conveys is generally that the initial temperatures make a difference in the radius evolution. The internal energy of the hot initial planet is below the gravitational binding energy of $\frac{3}{5} G M^2 / R$.} The dashed gray line shows the fractional differences between the radii of hot models and the cold models. All models shown in Figure~\ref{fig:fig_3_10_evol} assume a homogeneously mixed envelope with $Z = 0.5$, corresponding to a mean molecular weight of 4.1 amu, where the envelope is 5\% of their total mass. The hotter model (blue) retains a larger radius than the colder model (red) by 10\% at 100 Myrs and by 5--7\% even after 5 Gyrs. By 10 Gyrs, the radii of the hot and cold models differ by only 0.1\%. The miscibility curves of \citet{StixrudeGilmore2025} are shown in green, showing that MgSiO$_3$ remains miscible with hydrogen above these temperatures. The envelopes sustain a higher temperature than most of the hydrogen-water miscibility curve of \cite{Gupta2025}. We note that the cooling of the envelope is partially controlled by the assumed equilibrium temperatures and the metal mass fraction of the envelope. We have used the same 400 K equilibrium temperature, corresponding to $\sim$10 times the solar flux incident on Earth. The mantles of these models spend most of their evolution above the silicate melting curves of \cite{Deng2023}, and the cores do not drop below the melting curves of Fe-alloys Fe$_{16}$Si \citep{Fischer2012} or Fe$_{17}$Si \citep{Ezenwa2024} that have been argued to dominate the core composition of planets such as the Earth. However, the cores do intercept the pure-Fe melting curves of \cite{Zhang2015} and \cite{Gonzalez-Cataldo2023} (not shown) at ages between 2 and 5 Gyr.

We parameterize the Fe$_{16}$Si melting data from Figure~1 of \citet{Fischer2012}
by fitting it to the Simon--Glatzel relation, obtaining
\begin{equation}
T_m(P) = 1500 \left[ 1 + \frac{P - 2.0}{11.29} \right]^{0.36},
\end{equation}
where $T_m$ is in K and $P$ is in GPa. For Fe$_{17}$Si, we use the Simon--Glatzel coefficients presented by \citet{Ezenwa2024}.

\begin{figure*}[ht!]
\centering
\includegraphics[width=0.9\textwidth]{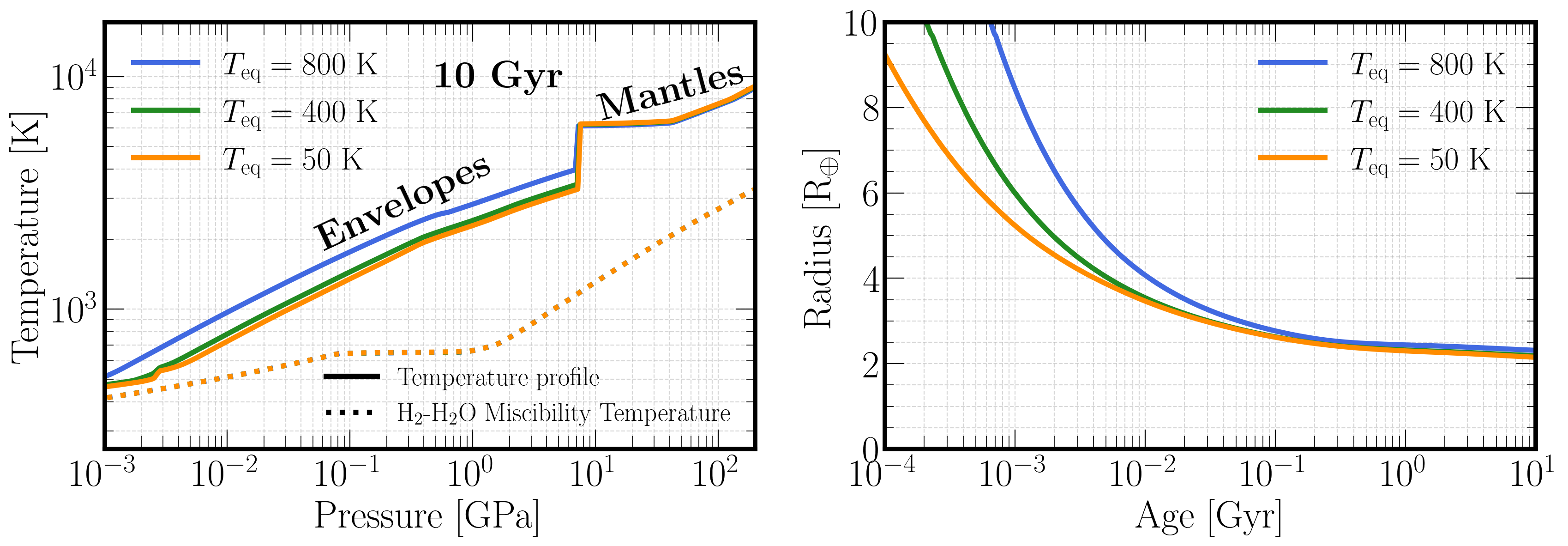}
\includegraphics[width=0.9\textwidth]{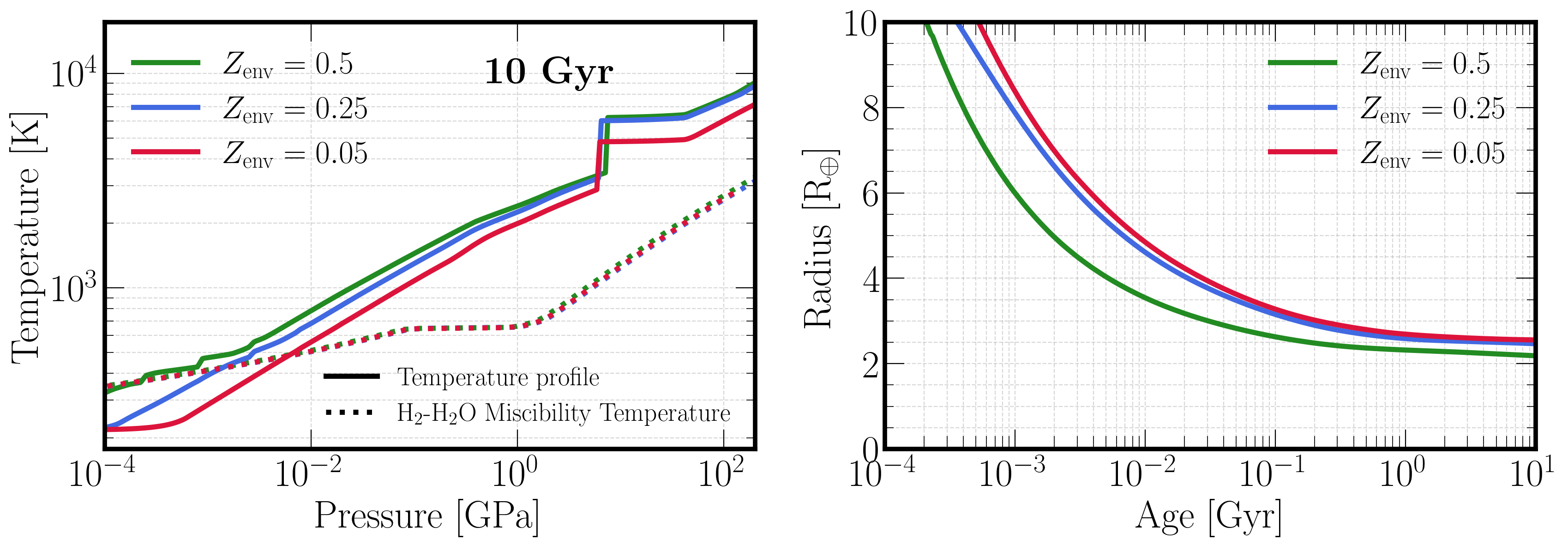}
\includegraphics[width=0.9\textwidth]{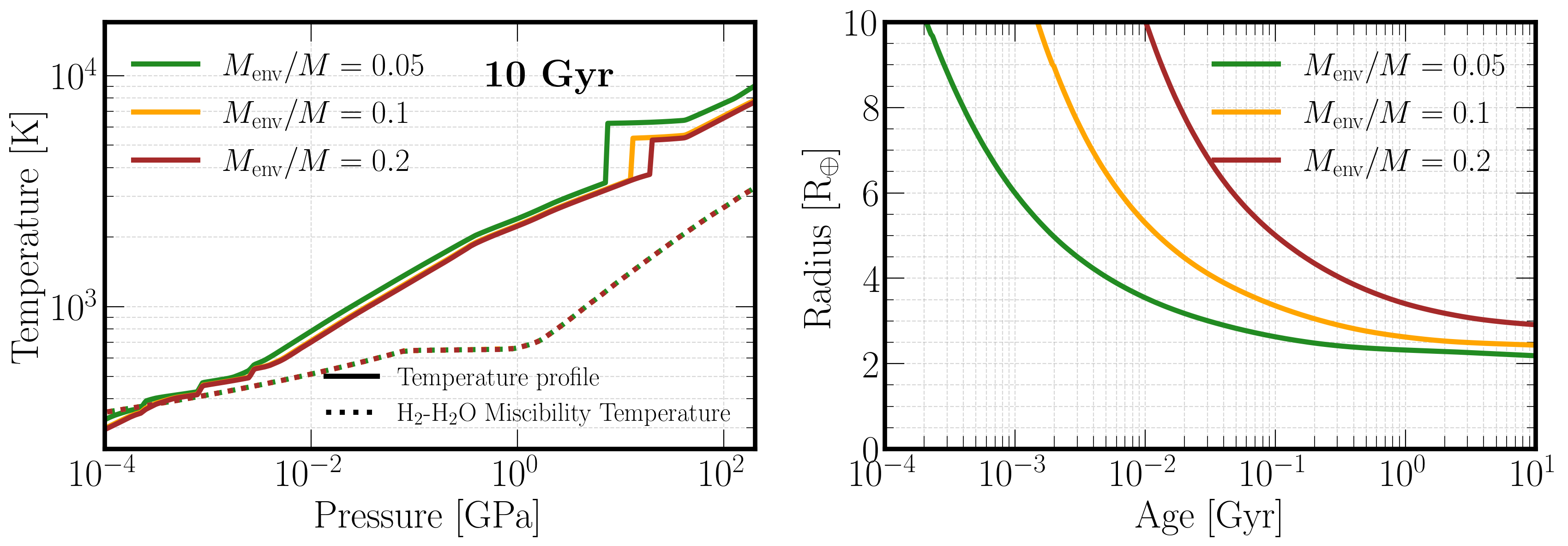}
\caption{The same 3 \mearth model shown in Figure~\ref{fig:fig_3_10_evol}, shown in green throughout this figure, is compared here with different equilibrium temperatures (top), different envelope metal mass fractions (middle), and different envelope mass sizes (bottom). The water-hydrogen phase separation curves of \cite{Gupta2025} are plotted for each model as dotted lines.  In the top row, three models with equilibrium temperatures of 800 K, 400 K, and 50 K are compared at their final states in the left panel. The right panels show the radius dependence on the equilibrium temperature. The dependence on the envelope metal mass fraction ($Z_{\rm env}$) is shown in the middle row. Lower envelope metal fractions can access the water-hydrogen phase-separation curves due to enhanced cooling, but only at pressures $\sim$10--50 bar. The bottom row compares the same model in green with similar models with different envelope mass fractions. Larger envelope fractions of 10\% and 20\% lead to significantly larger radii at early ages, and intercept the water-hydrogen miscibility curve at the same temperatures and pressures as the green model. }
\label{fig:fig_3_teq}
\end{figure*}

The evolution of the envelope also depends on the equilibrium temperature (i.e., the distance from its star) as well as the metallicity of the envelope. The top row of Figure~\ref{fig:fig_3_teq} shows the final temperature profiles of the same 3 \mearth model shown on the left column of Figure~\ref{fig:fig_3_10_evol} compared to identical models at 800 K and 50 K equilibrium temperatures. The stellar luminosity evolution is not modeled here. From the top row of Figure~\ref{fig:fig_3_teq}, we find little dependence of the onset of water-hydrogen immiscibility on the equilibrium temperature. The dependence on metal mass fraction is shown in the middle row of Figure~\ref{fig:fig_3_teq} for the same 400 K equilibrium temperature, and the dependence on envelope mass is shown in the bottom row. While the radius evolution varies significantly with metallicity and envelope mass fraction (middle and bottom rows), we find no significant difference in the inner envelope temperatures. The outer envelopes of the middle-row models permit greater water advection, but only at lower water abundances. We recall that our atmosphere model depends on the local metal fractions (Section~\ref{subsec:atm}), allowing greater cooling in the outer regions. Hotter equilibrium temperatures keep envelopes hotter (as expected) and above the miscibility curves of \cite{Gupta2025}. In contrast, envelopes with lower metal mass fractions allow partial water-hydrogen phase separation due to enhanced cooling. 

\begin{figure}[ht!]
\centering
\includegraphics[width=0.43\textwidth]{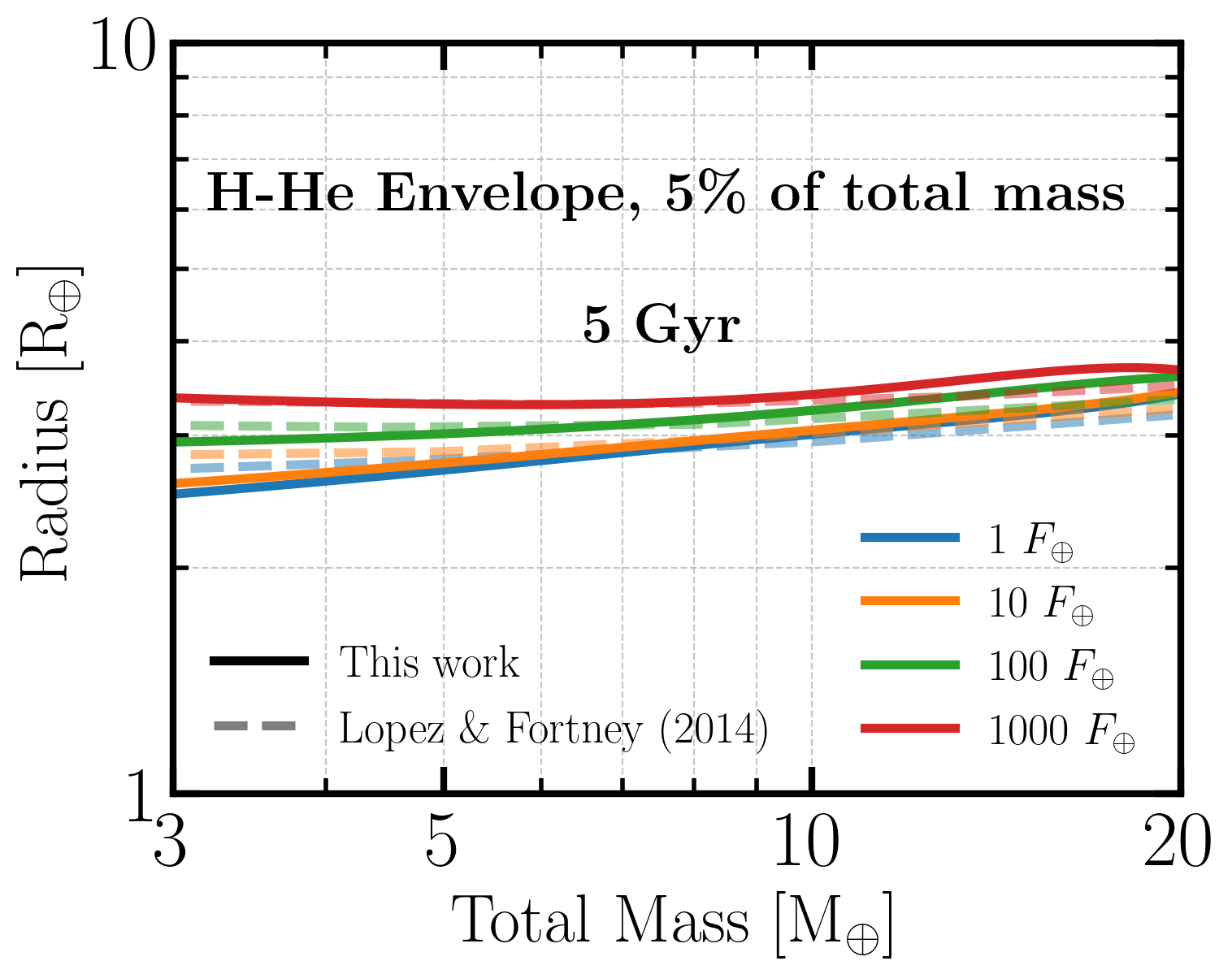}
\caption{Comparison of the sub-Neptune mass-radius relation obtained by \cite{Lopez2014} as a function of stellar flux (color lines).  \cite{Lopez2014} assumed isothermal interior temperature profiles, used the \cite{Saumon1995} H-He EOS, and an ANEOS olivine EOS \citep{Thompson1990} to represent the mantle composition. The differences in radii $\lesssim 5$ \mearth are attributed to a lower opacity in our envelopes, allowing for more cooling and contraction. This is observed in Figure 1 of \cite{Lopez2014}. There, they compare enhanced opacities and solar opacities, showing similar differences in radii. We obtain larger radii at higher masses due to our adiabatic mantle and core structure compared to the isothermal structures assumed by \cite{Lopez2014}. We use the H-He EOS of \cite{Chabrier2021}, an updated EOS for liquid MgSiO$_3$ \cite{Luo2025}, and assume an adiabatic interior temperature profile. }
\label{fig:fig_mr_relation}
\end{figure}

We compare our models to a part of the mass radius relation from \cite{Lopez2014} in Figure \ref{fig:fig_mr_relation}, where the color lines represent various incident fluxes, showing the radius at 5 Gyr as a function of total planet mass. We computed our own models (solid lines) within this parameter range to compare them with those of \cite{Lopez2014} (dashed curves). We obtain larger radii for larger planets because of adiabatic interior structures, compared with the isothermal structures assumed by \cite{Lopez2014}. Note that our models in Figure~\ref{fig:fig_mr_relation} do not have metals in the envelope, so their envelopes are purely H-He, comprising 5\% of the total mass of each planet. We find more pronounced differences for masses smaller than $\lesssim$5 M$_\oplus$, where the effects of initial temperatures are more significant. Since the envelopes of the models shown in Figure~\ref{fig:fig_mr_relation} are H-He only, the opacities are lower, allowing for more cooling and contraction. \cite{Lopez2014} used enhanced opacities at 50 times solar metallicity and 1 times solar metallicity in their Figure 1. 

\subsection{Envelope-mantle Stratification} \label{subsec:stratification}

The envelopes of sub-Neptunes may not be homogeneous. Stable regions may arise at formation or due to immiscibility \citep{StixrudeGilmore2025, Rogers2025}. In this Section, we incorporate stably stratified regions in the envelopes of sub-Neptunes and demonstrate that such stable regions contract more slowly, largely due to less efficient heat transport. This effect enlarges radii by $\sim$5-7\%. 

\begin{figure*}[ht!]
\centering
\includegraphics[width=\textwidth]{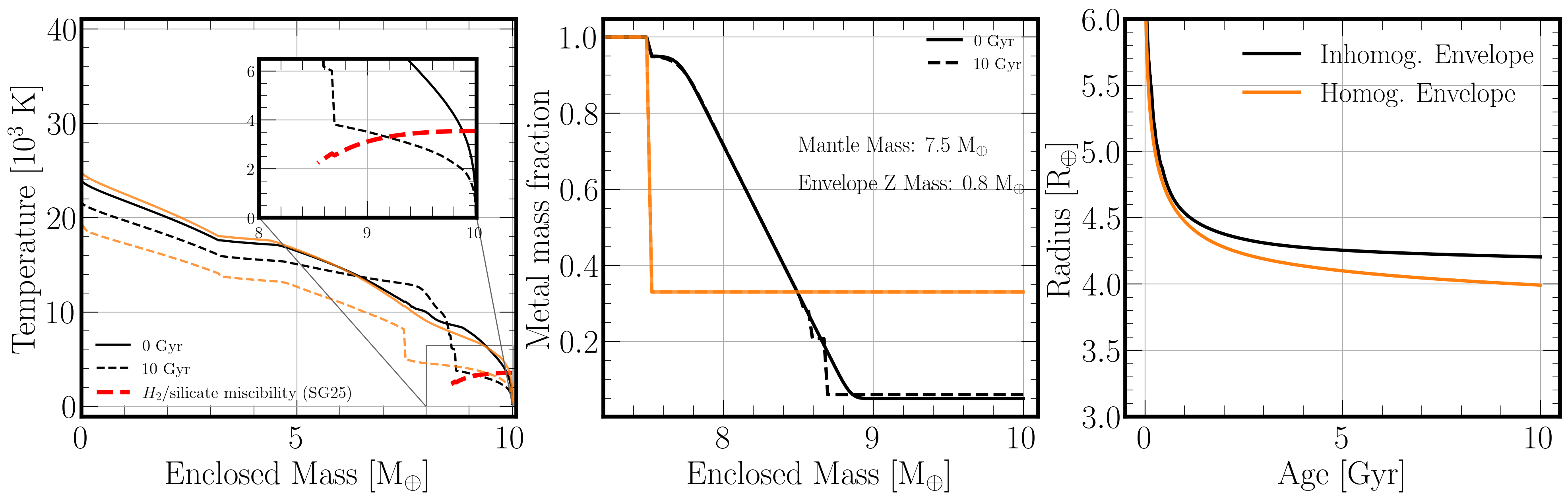}
\caption{Example evolution models of a 10 \mearth sub-Neptune structures with (black) and without (orange) extended mixed mantles, with 7.5 \mearth mantle mass. The total mass of heavy elements in the envelope is set to 0.8 \mearth for both planets. The left, center, and right columns show the evolution of the temperature profile, the metal mass fraction profile, and the radius, respectively, for both the stably stratified and homogeneous models. The thick red line in the left panel inset is the miscibility curve of \citet[][SG25]{StixrudeGilmore2025}, above which rock and H$_2$ can remain mixed. The heavy elements in regions below the red dashed line, or the silicate-hydrogen miscibility curve, are represented by water, and those above it are MgSiO$_3$. The homogenizing effects of convective mixing are included in the envelope. At late ages, contraction heats the stably stratified regions of the lower envelope. The effects of convective mixing are visible in the center panel, where the inhomogeneous model has accumulated more heavy elements but has not mixed its initially stable layers. Since the homogeneous model has only the EMB as a stabilizing molecular weight gradient, it contracts by $5.3\%$ more than the inhomogeneous model.}
\label{fig:fig_homog_vs_inhomog}
\end{figure*}

We show in Figure~\ref{fig:fig_homog_vs_inhomog} two example 10 M$_\oplus$ models harboring the same heavy element mass (8.3 M$_\oplus$). The mantle and core comprise 7.5 \mearth while the envelope harbors 0.8 \mearth of heavy elements. The difference between the black and yellow models in Figure~\ref{fig:fig_homog_vs_inhomog} is that the orange model has a homogeneous envelope, and the black model an inhomogeneous one. The inhomogeneous structure creates a stably stratified region due to its negative molecular-weight gradient. The effects of convective mixing are included. Both models start with a similar temperature profile. Still, the black model features a gradual, heavy-element profile in which MgSiO$_3$ is mixed with the hydrogen-helium mixture (heavier by 27\% by mass), providing a smooth transition between the H-He right envelope and the mantle. The thick red-dashed line in the left panel of Figure~\ref{fig:fig_homog_vs_inhomog} is the silicate-hydrogen miscibility temperature, above which H$_2$ and rock will be miscible \citep{StixrudeGilmore2025,Rogers2025}, justifying our 9 M$_\oplus$ placement of a gradually stably-stratified region. Both models predict different radii after 1 Gyr by $\sim$6\% (right panel), in approximate agreement with \cite{Rogers2025}. This difference grows with lower planet masses and larger stable regions.

\begin{figure}[ht!]
\centering
\includegraphics[width=0.45\textwidth]
{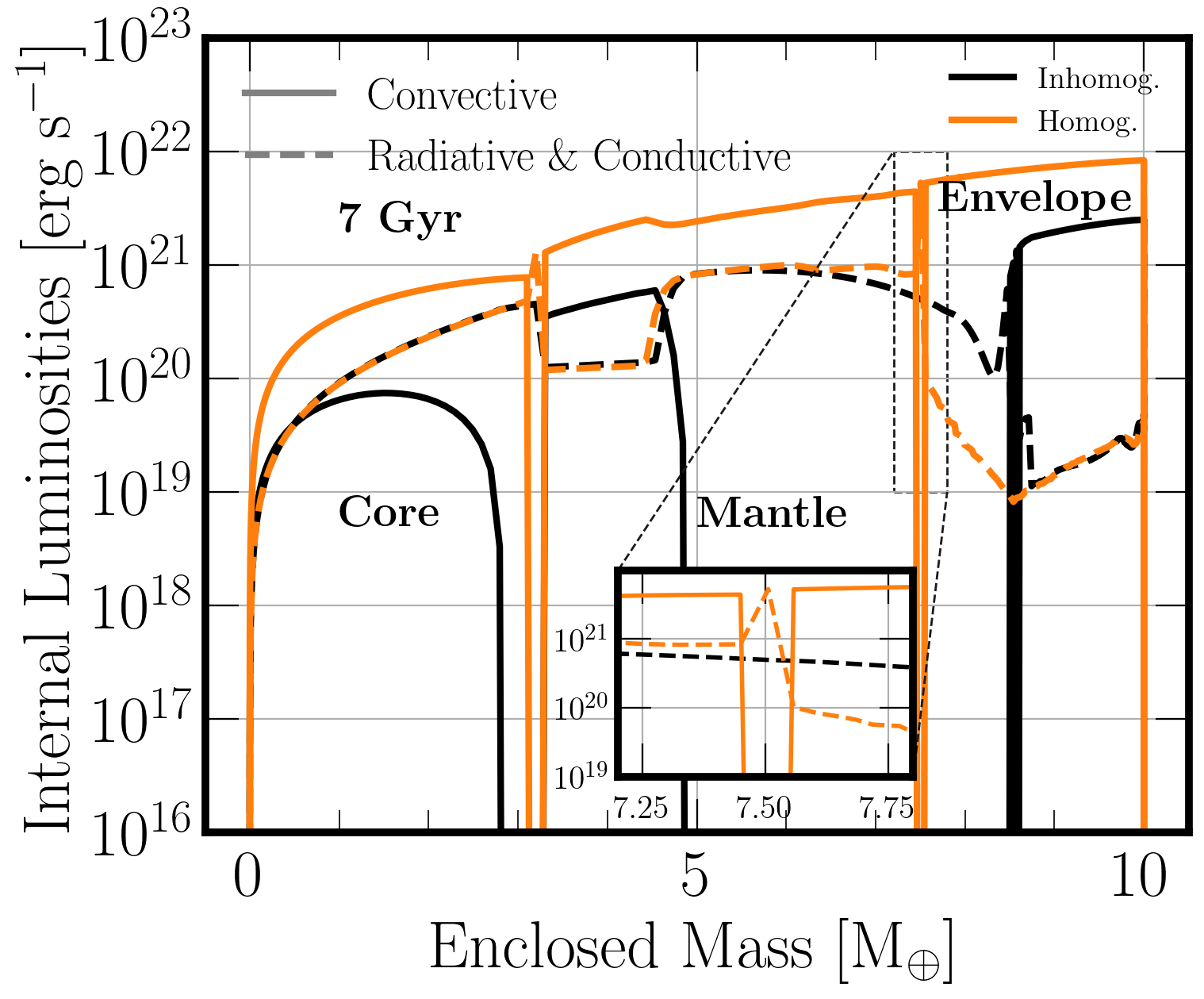}
\caption{Convective and combined radiative and conductive internal luminosities (or ``fluxes'') of the 10 M$_\oplus$ models shown in Figure~\ref{fig:fig_homog_vs_inhomog} at 7 Gyr as a function of enclosed mass coordinates. The orange and black models are homogeneous and stably stratified, respectively. The large stratified region of the black model transports heat solely by conduction and radiation, with no convective flux in the outer homogeneous envelope. The luminosities across the EMBs of the models are shown as insets in each panel. Convective heat transport is inhibited in the upper mantle of the inhomogeneous model due to reduced heat transport to the outer regions, resulting in a shallower temperature profile. This is an expected behavior if the interior heat cannot be dissipated. Within these boundaries in the homogeneous model, the convective luminosity drops to zero due to the steep molecular-weight and entropy gradients, after which non-convective luminosities take over. While the non-convective luminosities are continuous, the composition gradient barrier halts convective thermal \textit{and} composition flux, and this lack of continuous composition flux inhibits mantle cooling. This difference in the heat flux between a stably-stratified interior and an adiabatic interior leads to the radius differences shown in the right panel of Figure~\ref{fig:fig_homog_vs_inhomog}.}
\label{fig:homog_inhomog_lums}
\end{figure}

The larger radii at evolutionary timescales are explained by higher interior temperatures in the stably stratified model due to the lack of convective heat transport. Figure~\ref{fig:homog_inhomog_lums} shows an example of the convective, radiative, and conductive internal luminosities as a function of mass shell boundaries of the entire structures shown in Figure~\ref{fig:fig_homog_vs_inhomog} at an age of 7 Gyr. The convective luminosities of the homogeneous model (orange) exceed those of the stably-stratified model (black), thereby enabling more efficient cooling. The envelope convective region of the stable model (black) is evident through an increase in convective luminosity in the outer 10\% of the planet. The EMB of the homogeneous model is characterized by a sudden decrease in convective luminosity at 0.75 M$_P$, where conduction is the sole heat luminosity. The inset in Figure~\ref{fig:homog_inhomog_lums} highlights the EMB, showing a peak of non-convective flux at this transition layer. The inset shows a peak in the conductive and radiative luminosities at the boundary transition. This is due to the outgoing luminosity from a given mass shell being the same luminosity received by the next cell, and these are calculated at the cell boundaries, or ``faces'' \citep[See Figure 10 in][]{Sur2024a}. Thus, the luminosities of the outermost mantle cell boundary are transferred to the inner cell boundary of the EMB. This peak is not visible in the homogeneous model (in black) since the luminosities in the upper parts of its mantle are already non-convective. As described in Section~\ref{subsec:entropy}, under the MLT methodology, heat is transported by thermal fluid parcels that disperse their heat into the surrounding fluid. The inefficient cooling of the envelope is therefore primarily driven by a lack of composition flux, which is inherent in convective regions (Eq.~\ref{eq:conv_mixing}). This causes the steep temperature gradients observed on the left panel of Figure~\ref{fig:fig_homog_vs_inhomog}.

\subsection{Evolution of Silicate Rain} \label{subsec:miscibility}

\begin{figure*}[ht!]
\centering
\includegraphics[width=\textwidth]{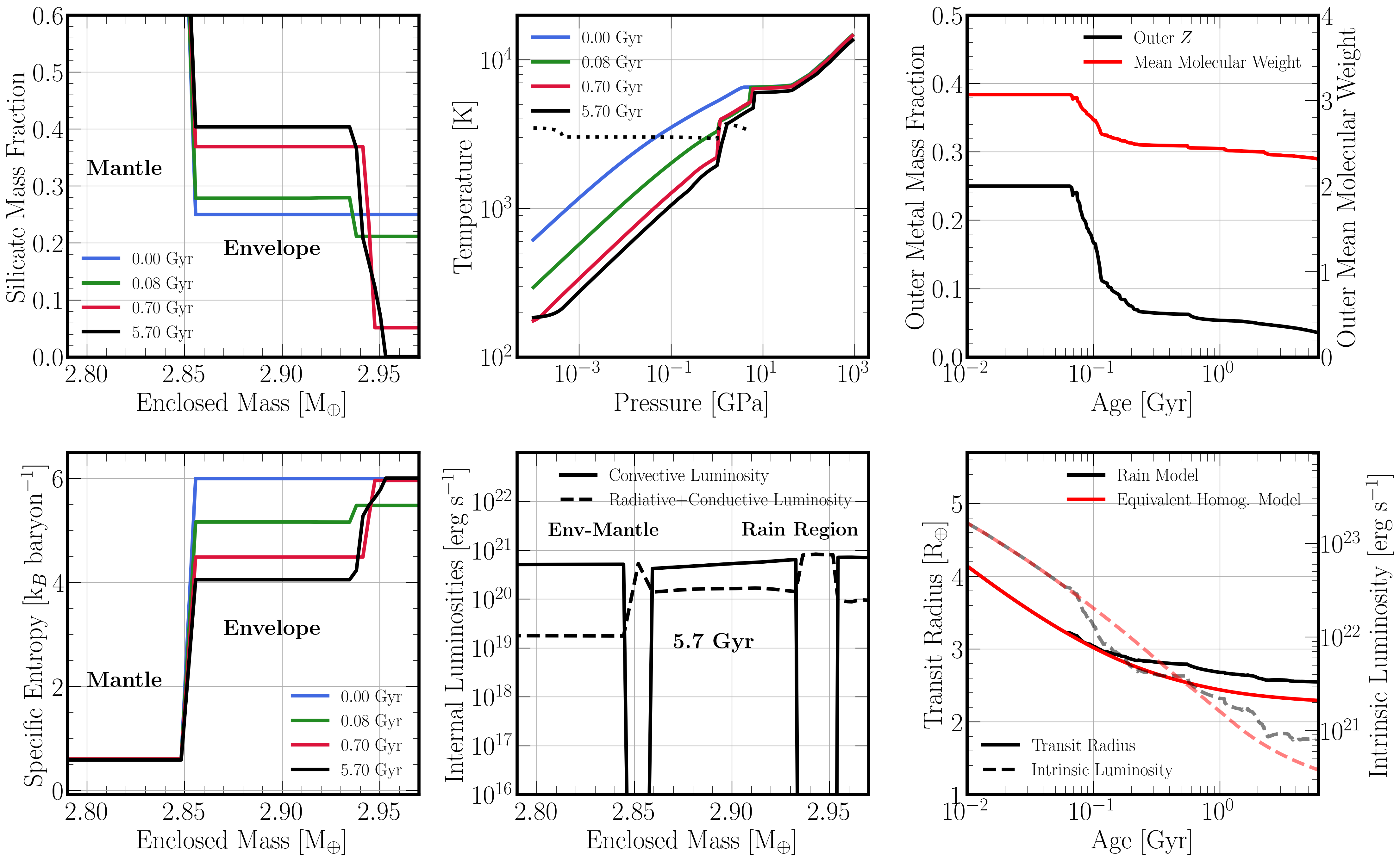}
\caption{Silicate phase separation regions form stable regions in the envelope of a 3 M$_\oplus$ sub-Neptune example model. The silicate rain region is determined by the interception of the temperature profiles (top center) with the miscibility curves of \cite{StixrudeGilmore2025}, shown as dotted lines in the top center panel. We calculate the miscibility temperatures and equilibrium abundances along the temperature-pressure profiles of the envelope. The top left panel shows the evolution of the silicate (i.e., MgSiO$_3$) mass fraction profile, and the top right panel shows the evolution of the outer metal abundance and mean molecular weight (black and red, respectively). Solid color lines indicate different ages. The bottom row shows the evolution of the entropy profile. The bottom-center panel shows a snapshot of the internal luminosities/fluxes at 5.7 Gyr, highlighting the stable regions of the EMB and the rain region. The rain model begins homogeneously mixed (blue lines) with a liquid MgSiO$_3$ mass fraction of 0.25 and evolves to deplete its initially mixed liquid MgSiO$_3$ over Gyr timescales. This process is not instantaneous due to the modeling methods outlined in Section~\ref{sec:methods}. The advection of silicates from the outer regions to the inner regions creates a stable layer above the already convectively stable envelope-mantle boundary layer, affecting the evolution of the radius by an additional $\sim$5\%. The bottom-right panel shows the evolution of the transit radius (solid lines) for the rain model, compared to an equivalent homogeneous model shown in red and black, respectively. We show the intrinsic luminosity evolution using faint dashed lines in the bottom-right panel. The equilibrium temperature is cold (50 K) to illustrate enhanced depletion, since prolonged cooling drives immiscibility. Similar models at 800 K deplete their outer envelopes for longer (See Figure~\ref{fig:fig_misc_3mearth_50Kvs800K}).}
\label{fig:fig_misc_3mearth}
\end{figure*}

In this Section, we introduce models with liquid silicate (MgSiO$_3$) phase separation as an evolutionary process that affects the radius at ages $\gtrsim 0.1$ Gyr by $\sim$5\%, depending on the envelope silicate abundance. An example 3 M$_\oplus$ evolution model is shown in Figure~\ref{fig:fig_misc_3mearth} with an equilibrium temperature of 50 K. This model undergoes phase separation of silicates with hydrogen and helium (with a helium mass fraction relative to hydrogen of 0.05). The model in Figure~\ref {fig:fig_misc_3mearth} has an envelope mass that is 5\% of its total mass, while 95\% of its mass is in its mantle and core. Silicates account for all the metal mass in the envelope. The liquid silicate metal fractions distributed across the envelopes of models in this section are described by the same liquid EOS as in the mantle \citep{Luo2025}. The top row of Figure~\ref{fig:fig_misc_3mearth} illustrates the depletion of silicates from the outer layers (top left). Here, a larger-radius (by $\sim$5\%) model is compared with a homogeneous model with the same envelope metal and total mass (bottom right). This difference is due to the convectively stable silicate rain layer in the former and the gravitational energy release from silicate rain that heats up the outer regions, shown in the lower left panel of Figure~\ref{fig:fig_misc_3mearth}. The rain model also exhibits an intrinsic luminosity twice that of the homogeneous model (bottom right).

An illustration of the initial and final conditions of the model shown in Figure~\ref{fig:fig_misc_3mearth} is shown in Figure~\ref{fig:fig_misc_drawing}. The silicates in the envelope phase-separate from the hydrogen and helium components in regions above the EMB, thereby bifurcating the envelope into a silicate-poor upper layer and a silicate-rich lower layer. This evolutionary process decreases the metallicity of the atmosphere over time and thus reduces the mean molecular weight of the atmosphere, as shown in the upper right panel of Figure~\ref{fig:fig_misc_3mearth}. The temperatures of the mantle and core cool only modestly, as shown in the upper center panel of Figure~\ref{fig:fig_misc_3mearth}. 

The top row of Figure~\ref{fig:fig_misc_3mearth_50Kvs800K} shows the same model as Figure~\ref{fig:fig_misc_3mearth} compared to an identical model with an equilibrium temperature of 800 K. This higher equilibrium temperature inhibits silicate rainout by raising envelope temperatures and increasing convective mixing, which homogenize compositional gradients. Models with larger envelopes, which constitute 30\% of their mass rather than 5\%, exhibit silicate immiscibility at the same pressures and temperatures. This is shown in the bottom row, middle panel of Figure~\ref{fig:fig_misc_3mearth_50Kvs800K}. Since these pressures lie deeper in the structure, the rain location shifts inwards, as illustrated in the left panel. Even with larger envelopes, the silicate rain region bifurcates the envelope into a hydrogen- and helium-rich upper layer and a silicate-rich inner layer. As the bottom row of Figure~\ref{fig:fig_misc_3mearth_50Kvs800K} also shows, in a more massive envelope (dashed line), a deeper silicate rain region is located below more mass, so the rain depletion is slower than in the thin envelope case (solid). Together, Figures~\ref{fig:fig_misc_3mearth} and \ref{fig:fig_misc_3mearth_50Kvs800K} show that: 1) Silicate depletion could occur over the evolution of sub-Neptune interiors, 2) silicate rain can further increase the radius at late ages, thereby decreasing their mean densities, 3) that cold, long-period sub-Neptunes could experience significantly more phase separation than hot, short-period sub-Neptunes, and 4) that the envelope size affects the depletion rates at the same temperatures and pressures. This is discussed in Section~\ref{sec:discussion}.

\subsection{Example Models: GJ1214 b, K2-18 b, TOI-270 d, \& TOI-1801 b} \label{subsec:case_studies}

In this Section, we aim to demonstrate that inefficient mantle cooling with hot starts and small envelopes is sufficient to explain the mean densities of sub-Neptune exoplanets currently hypothesized to be water worlds. We explore the interior evolution of four exoplanets: GJ 1214 b \citep{Charbonneau2009}, K2-18 b \citep{Monet2015}, TOI-270 d \citep{VanEylen2021}, and TOI-1801 b \citep{Mallorquin2023}.\footnote{The reader should note that these models are not ``fits,'' but rather demonstrations.} Since the cooling of the mantle dominates the thermal evolution, we forego inhomogeneous evolution (demonstrated in Sections~\ref{subsec:stratification} and \ref{subsec:miscibility}). Given the uncertainties surrounding inhomogeneous evolution, even for the Solar System gas giants \citep{Vazan2018, Tejada2025, Sur2025a}, we refrain from proposing inhomogeneous evolution models for specific sub-Neptune exoplanet observations. As such, we proceed with the assumptions made in Section~\ref{subsec:entropy}, where only the EMB limits the rate of cooling of the mantle and core. More detailed calculations of these sub-Neptunes will be presented in future work. The mass, radius, equilibrium temperature, age, and mean density measurements of each exoplanet are tabulated in Table~\ref{tab:subneps}. These exoplanets were chosen to cover a wide range of ages. Based on stellar spot detections, GJ 1214 is estimated to be between 6 and 10 Gyrs old \citep{Mallonn2018}, K2-18's gyrochronology estimates place it at $2.4 \pm 0.4$ Gyrs old \citep{Guinan2019}, and kinematic analysis of TOI-1801 places its age between 600 and 800 Myrs \citep{Mallorquin2023}. The age of TOI-270 is currently unconstrained, so we use a wide age range of 1-10 Gyr to calibrate its evolution model. 

For these demonstrations, we model each planet using the mean of each measurement, so we take 8.17, 8.63, 4.78, and 5.74 M$_\oplus$ for GJ 1214 b, K2-18 b, TOI-270 d, and TOI-1801 b, respectively. We take a similar approach to calibrate the remaining observables for each model. We use a homogeneous envelope metal abundance of $Z=0.67$, represented by water, by mass (mean molecular weight of 5.56 amu), corresponding to $\sim100$ times the solar metallicity for GJ 1214 b \citep{Miller-Ricci2010, Nixon2024}. We do not explore the effects of the envelope metallicity here, so we retain the envelope metal mass fraction across all four planets. The reader should note that although metallicity affects the evolution of the radius, we aim to show that hot mantles are sufficient to explain these exoplanet radii. A metallicity of 100 times the solar value is already high, so we chose this abundance value to mitigate the effects of larger radii at lower metal fractions. Lower metal mass fractions lead to larger radii, so to account for the observed radii, even larger mantle and core sizes are required. This illustrates our point that, at either low or high metallicities, large liquid mantles can explain the observed radii. 

We find that combined mantle and core masses of 7.05, 7.95, 4.54, and 5.5 M$_\oplus$ produce the observed radii and mean densities at the relevant ages for each respective planet. The rocky interiors thus comprise 87.5\%, 92.1\%, 95\%, and 96\% of the total mass of GJ 1214 b, K2-18 b, TOI-270 d, and TOI-1801 b, respectively.  The GJ 1204 b and K2-18 b models keep a 1:2 mass ratio between the core and the mantle. To better match the small radii of TOI-270 d and TOI-1801 b, the iron-rich core mass is higher at 3 and 2 M$_\oplus$, respectively. The initial mantle temperatures of the TOI-1801 were selected to be 4000 K cooler than GJ 1214 b and K2-18 b to match the radius better (this is discussed in Section~\ref{sec:discussion}).  The measured equilibrium temperatures shown in Table~\ref{tab:subneps} were used to simulate the effects of stellar irradiation corresponding to each planet, which are already included in our atmosphere models (Section~\ref{subsec:atm}). The evolution of each model matching the observed parameters is shown in Figure~\ref{fig:fig_case_studies}. These models are carried out to 10 Gyr, and their transit radii are calculated using Eq.~\ref{eq:deltaz_guillot}, shown as dashed lines in the left panel of Figure~\ref{fig:fig_case_studies}. The temperature profiles shown in the right panel are their initial conditions (solid lines) and their estimated present-day temperatures (dashed). The TOI-1801 b model did not cool much due to its young (700 Myr) estimated age. The example interior models of GJ 1214 b, K2-18 b, TOI-270 d, and TOI-1801 b shown in Figure~\ref{fig:fig_case_studies} show that a liquid rocky mantle and core can harbor sufficiently low densities to account for the observed radius and mean densities of these sub-Neptunes. This is possible because their heat is dissipated inefficiently through the processes outlined in Section~\ref{subsec:entropy}.

\section{Discussion}\label{sec:discussion}

In the Solar System, inhomogeneous compositional profiles and the resulting regions stable against convection appear ubiquitous. Prominent examples include the ``fuzzy'' core of Jupiter inferred from \textit{Juno} gravity measurements \citep[e.g.,][]{Wahl2017, Militzer2024}, the stable region inside Saturn hosting gravity modes detected via ring seismology \citep[e.g.,][]{Fuller2014b, Mankovich2021}, and the inhomogeneous, non-adiabatic interiors invoked for Uranus and Neptune to explain their anomalous luminosities \citep[e.g.,][]{Nettelmann2016}. It is increasingly evident that inhomogeneous formation and evolution models are required to reproduce the present-day properties of Solar System giants \citep{helledstevenson2017, Vazan2018, VazanHelled2020, Tejada2025, Sur2025a, Tejada2025b}. A primary thermal consequence of internal compositional gradients is the suppression of convective heat transport. These compositional gradients create thermal bottlenecks, causing the outer regions to cool while the deep interiors remain hot due to inefficient diffusion through stably stratified layers \citep{Tejada2025}. These Solar System lessons suggest that compositional gradients and non-adiabatic evolution could likely play a similar, critical role in shaping the radii and luminosities of sub-Neptunes.

\subsection{Implications of Inefficient Cooling across the EMB}\label{subsec:entropy_implications}

The consequences of inefficient heat transport across the EMB directly affect how sub-Neptune bulk compositions are inferred from mean densities. Mean densities from measured masses and radii are commonly used as proxies for bulk composition and to divide small planets into rocky and water-rich (including the proposed ``water-worlds'' or ``Hycean'') classes \citep[][]{Madhusudhan2021, Luque2022}. However, mass–radius data alone are highly degenerate with respect to interior structure \citep{RogersSeager2010,Dorn2015,Vazan2017}, and recent work has shown that the densities of putative water worlds can be reproduced by sub-Neptunes with relatively thin, H-He envelopes \citep{Rogers2023}. Independent statistical analyses likewise find no robust evidence for a distinct water-world population in current samples \citep{Dainese2025}. Recent studies on the interactions between hydrogen atmospheres and the interiors of sub-Neptunes indicate that most sub-Neptunes may harbor a liquid-rock interior \citep{Calder2025}. 

The work presented here corroborates the hypothesis that the ``water-world" class of sub-Neptunes \citep[e.g.,][]{Luque2022}, or in general, the sub-Neptune and super-Earth populations, are predominantly compatible with being part of the same parent population born with rocky interiors. Some sub-Neptunes may undergo total atmospheric loss \citep{Owen2013, Owen2016, Owen2017a, Ginzburg2016, Gupta2019, Gupta2020a, Owen2024}, cool more without an envelope, and become part of the super-Earth population. Others may retain a thin envelope, keeping their interior mantles hot and liquid, and undergo phase-separation processes, thereby explaining their densities, rather than necessarily being composed entirely of water and other ices. The recent detections of water vapor and methane on K2-18 b \citep{Benneke2019}, TOI-270 d \citep{Benneke2024}, and similar sub-Neptunes remain compatible with this interpretation, since high atmospheric volatiles can exist in the thin envelopes of these sub-Neptunes. Indeed, recent chemical models by \citet{Nixon2025} of atmosphere–magma-ocean interactions in TOI-270 d could explain the high abundances of water, methane, and carbon dioxide inferred by \citet{Benneke2024}.

Core-accretion models of super-Earths and sub-Neptunes naturally predict high temperatures at the base of accreted envelopes, reaching $\sim$10$^4$ K for typical accretion rates \citep{Lee2014, Ginzburg2016, Lee2018, Vazan2024}. The initial temperatures adopted in our hot-mantle models (Section~\ref{subsec:entropy}) therefore fall well within the range expected from formation theory. If sub-Neptune silicate mantles retain much of their formation heat over evolutionary timescales, then evolution models may help infer their formation entropies. This approach provides a window on initial conditions without the uncertainties inherent to formation calculations, and can therefore inform formation theory. Similar exercises exploring initial composition and entropy profiles have been carried out for Jupiter and Saturn \citep{Tejada2025, Sur2025a, Knierim2025b} and for Uranus and Neptune \citep{Tejada2025b}. Even under the limitations of exoplanet observations and interior modeling, extending this strategy to sub-Neptunes is a natural next step. \citet{Owen2020} showed that formation entropies can be constrained with sufficiently precise mass, radius, and age measurements, approaching the problem via photoevaporation limits on the minimum H–He mass that could have been lost while retaining the observed atmosphere. Our work suggests a complementary route: When accounting for the metal envelope fraction, at a given age, mass, and radius of a young sub-Neptune, only a subset of post-formation entropies yields temperatures hot enough to match the observations, as colder entropies yield denser interiors and smaller radii. We highlight here that these initial conditions are necessarily post-formation. Mass-loss processes during formation, such as ``boil-off'' \citep{Owen2016,Ginzburg2016}, may occur, but these models are initialized after such processes. This is further discussed in Section~\ref{subsec:caveats}. 

We apply this post-formation entropy approach to TOI-1801 b in Figure~\ref{fig:fig_case_studies}. Rather than increasing the mantle mass above 96\% or the core mass above 3 M$_\oplus$, we adopt a slightly lower specific mantle entropy of 0.6 k$_B$ baryon$^{-1}$ compared to 0.7--0.74 k$_B$ baryon$^{-1}$ for the other case studies. We emphasize here that studying the possible range of primordial entropies is better captured by stellar evolution-like codes, such as \texttt{APPLE} and \texttt{MESA}, since they can model inefficient mantle cooling without assuming that mantles cool at the same rate as envelopes and atmospheres. Moreover, the metal content of the envelope also affects the radius evolution of the planet, as shown in Figure~\ref{fig:fig_3_teq}. Metal-poor envelopes may appear larger than metal-rich envelopes by approximately 15--20\% at large ages (bottom right panel of Figure~\ref{fig:fig_3_teq}), and it is yet unclear which quantity is more dominant in the radius evolution of sub-Neptunes. An exploration of the mantle entropy and envelope metal abundance is strongly warranted in future work.  Improved measurements of radii, masses, ages, and atmospheric abundances will be essential for a more comprehensive census of post-formation entropies and temperatures.

\subsection{Implications of Silicate Rain}\label{subsec:silicate_implications}

Silicate depletion of the outer layers could heat the outer envelopes of sub-Neptunes, potentially inflating their radii by $\sim$5\% or more depending on the mass, initial thermal state, and the metal abundance of the envelope. Observationally, the general phase separation process suggests that envelopes could appear enriched at early ages and then become depleted. Once depleted, radii will then appear larger due to an increase in entropy (i.e., energy deposit) in the outer envelope from the inner regions. Phase separation of silicate and other constituents could occur early in the post-formation phases, depleting the outer envelope faster than evolutionary timescales, leading to young hydrogen-rich atmospheres compared to those of their host stars. Indeed, the 3 \mearth model shown in Figure~\ref{fig:fig_misc_3mearth} shows severe silicate depletion even by 100 Myrs. Given the processes of inefficient mantle cooling and phase separation of silicates and other constituents, larger sub-Neptune radii are expected even at early ages. Subsequent work based on these findings will aim to more rigorously predict the inflation of sub-Neptune radii, thereby better connecting our models to exoplanet demographic trends. 

\begin{figure*}[ht!]
\centering
\includegraphics[width=\textwidth]{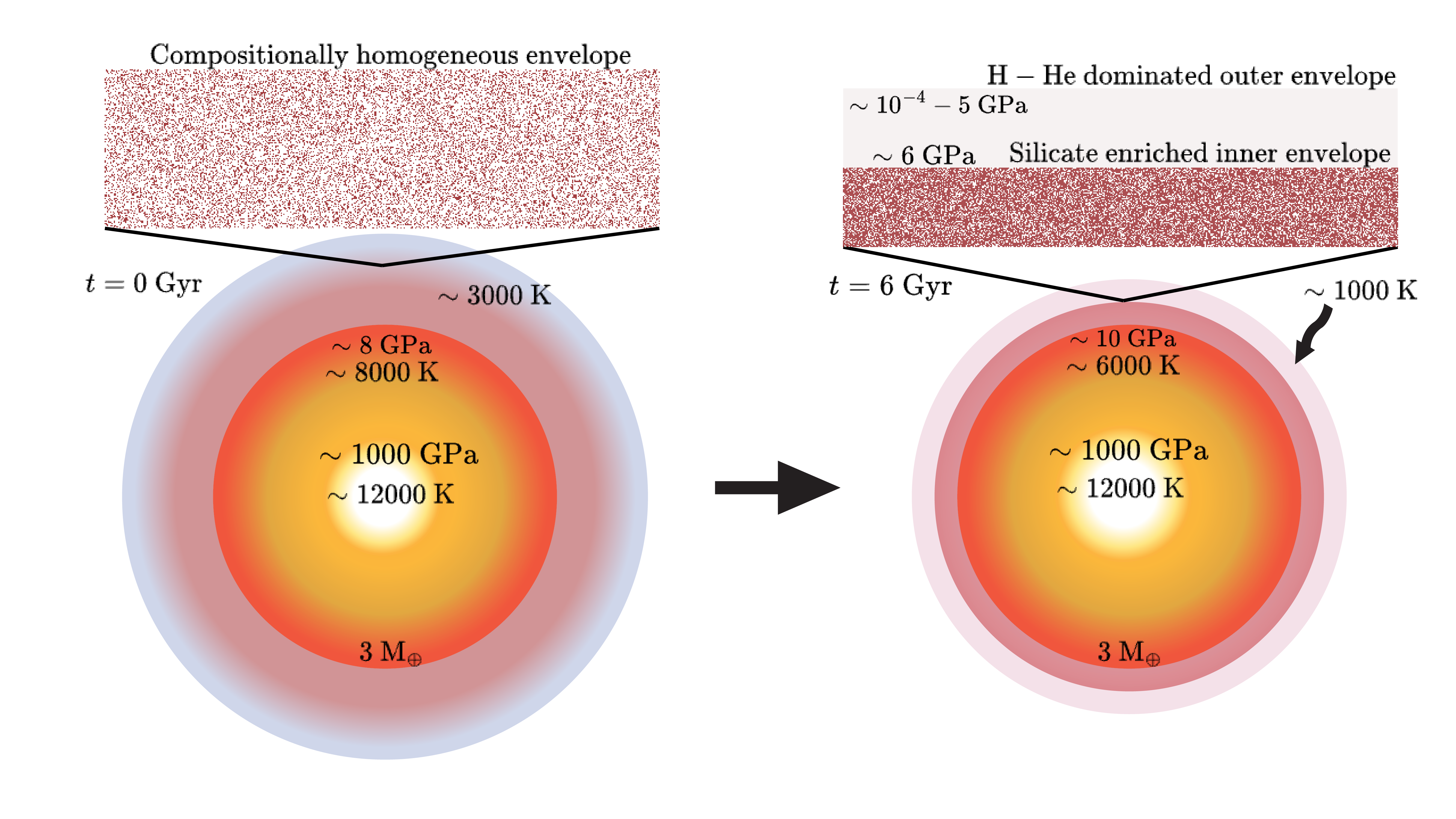}
\caption{Evolution illustration of the silicate phase separation layers depicted in Figure~\ref{fig:fig_misc_3mearth}. The left and right panels show rough sketches of the initial and final states of the 3 M$_\oplus$ model. This illustration, for illustrative purposes, shows approximately where the inner and outer envelopes partition to form a silicate-rich and a silicate-poor layer at 6 Gyr. The steep compositional gradients create a stably stratified zone between the inner and outer envelope convective regions. }
\label{fig:fig_misc_drawing}
\end{figure*}

\begin{figure*}[ht!]
\centering
\includegraphics[width=\textwidth]{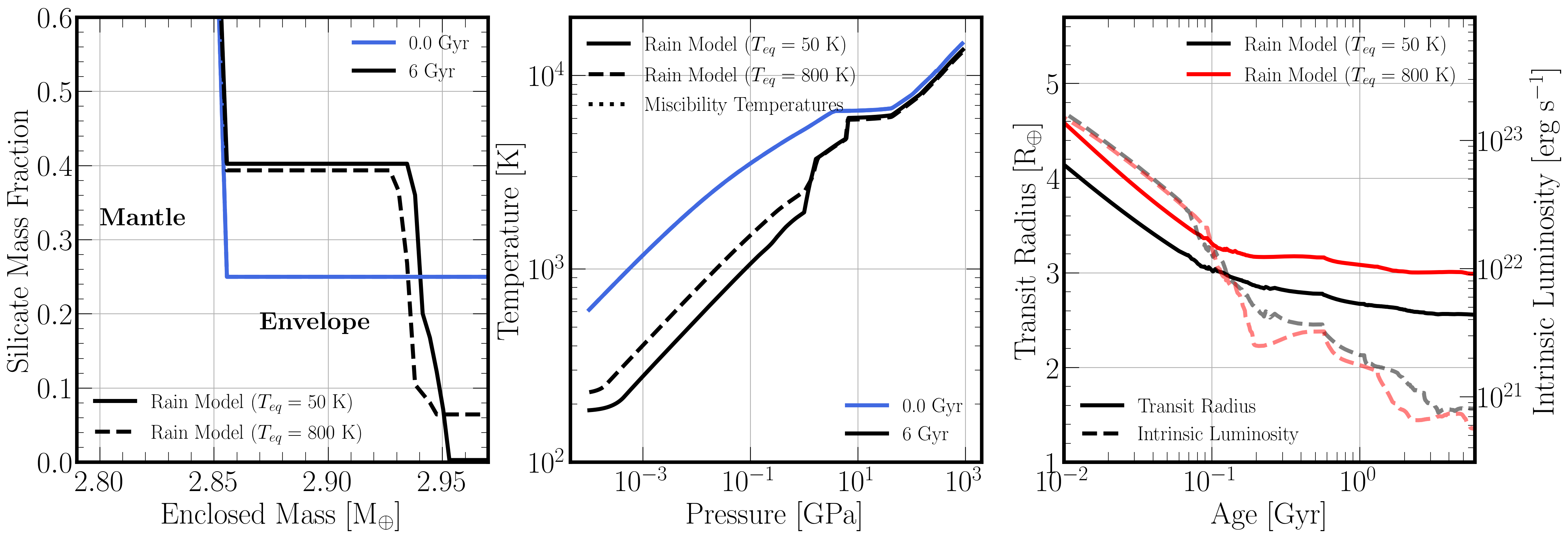}
\includegraphics[width=\textwidth]{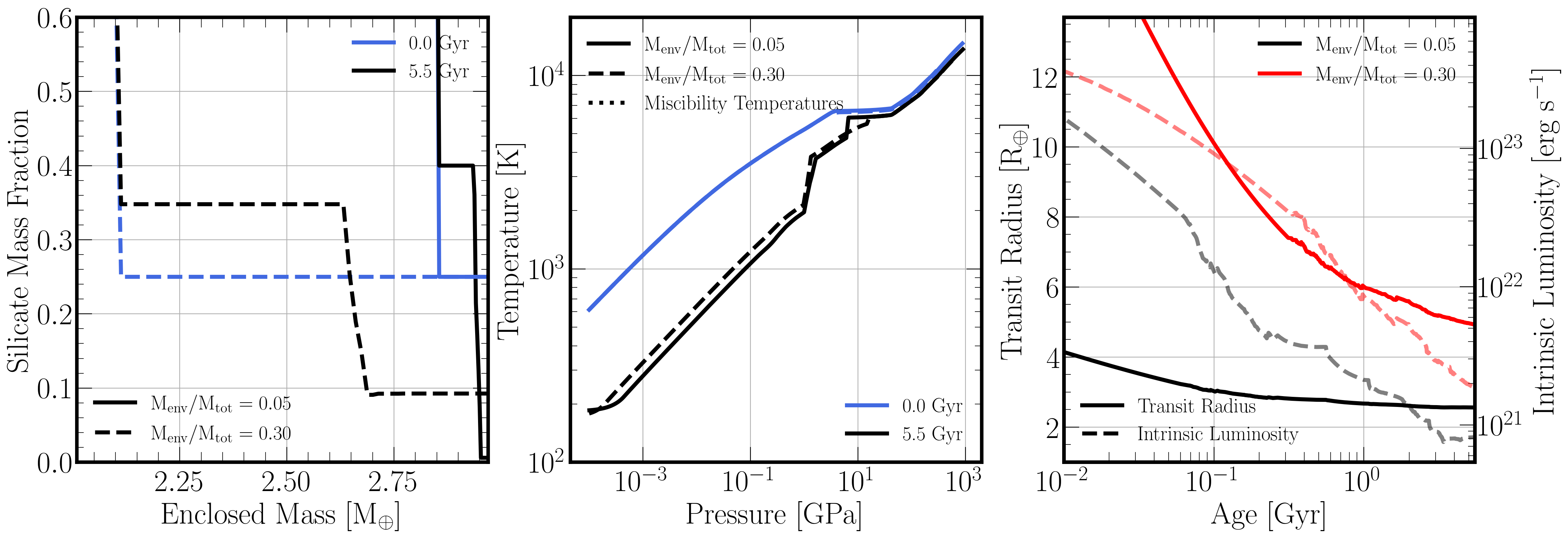}
\caption{Dependence of silicate depletion evolution on equilibrium temperature (top row) and envelope size (bottom row) for a 3 M$_\oplus$ model. Higher equilibrium temperatures and larger envelopes yield less depletion of the outer layers. Left/Center: Solid lines represent the fiducial model ($T_{\rm eq}=50$ K, $M_{\rm env}/M_{\rm tot}=0.05$) from Figure~\ref{fig:fig_misc_3mearth}. Blue lines denote the initial conditions, and black lines denote the evolved states. Right: Evolution of transit radius (solid) and intrinsic luminosity (dashed).
Top Row: The fiducial model is compared to a hotter version ($T_{\rm eq}=800$ K, dashed). Higher $T_{\rm eq}$ drives higher internal convective flux (right, red vs. black), which mitigates silicate rain (left), resulting in $\sim$9\% higher metal retention in the outer envelope.
Bottom Row: The fiducial model is compared to a large-envelope version ($M_{\rm env}/M_{\rm tot}=0.30$, dashed). While silicate rain occurs at similar pressures (center), the larger envelope retains more internal energy and mass above the rain layer, reducing the efficiency of advection-driven depletion. A larger envelope, therefore, sustains a silicate rain region above its EMB, just as a small envelope would.}
\label{fig:fig_misc_3mearth_50Kvs800K}
\end{figure*}

\begin{figure*}[t!]
\centering
\includegraphics[width=\textwidth]{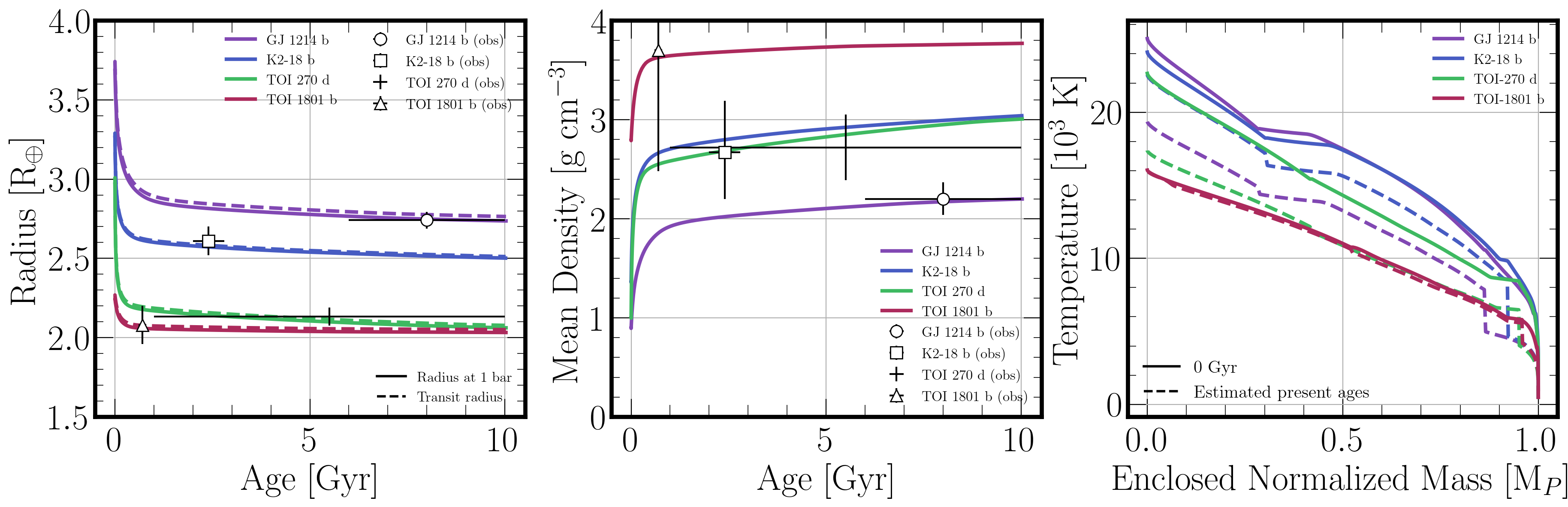}
\caption{Demonstrative evolution of models of GJ 1214 b \citep{Charbonneau2009}, K2-18 b \citep{Monet2015}, TOI-270 d \citep{VanEylen2021}, and TOI-1801 b \citep{Mallorquin2023}. The present-age radius of each planet can be matched (left panel) with hot interior MgSiO$_3$ liquid mantles, which remain hot, and at low enough densities (center panel) at their present age. The left panel shows the transit radius (in dashed lines) and the radius at 1 bar (in solid). The radius at 1 bar is our hydrostatic boundary condition, while the transit radius is calculated using Eqs~\ref{eq:rtransit} and \ref{eq:deltaz_guillot}. The right panel shows the initial temperature profile (solid lines), the median present age profile (dashed lines), and the final temperature profiles (dotted lines) at 10 Gyr. The median masses of 8.17 \citep{Cloutier2021}, 8.63 \citep{Cloutier2019}, and 5.74 M$_\oplus$ \citep{Mallorquin2023} were used for each respective exoplanet for these demonstrations. The white data icons show the measured radii at the estimated ages for each exoplanet. These values are tabulated in Table~\ref{tab:subneps}, where the mean values were used to calibrate each model.  The mantle masses are taken to be 7.05 and 7.95 M$_\oplus$ (87.5\% and 92.1\% of total mass, respectively) for GJ 1214 b and K2-18 b, and 4.54 and 5.5 M$_\oplus$ (96\% of total mass) for TOI-270 d and TOI-1801 b, respectively. The iron cores of GJ 1214 b and K2-18 b were maintained at a 1:2 total mass ratio with respect to the mantle, but this ratio was increased to include 2 and 3 M$_\oplus$ core masses for TOI-270 d and TOI-1801 b. We emphasize here that these models are not fits and should not be interpreted as definitive.}
\label{fig:fig_case_studies}
\end{figure*}

The silicate rain regions of the models shown in Section~\ref{subsec:miscibility} lie above the EMB regardless of their initial envelope abundance and their envelope sizes. The intersection between the miscibility temperatures and the envelope temperature profiles determines the location of the silicate rain layer. Since the miscibility temperatures decrease at pressures above $\sim$8 GPa, as shown in Figure~\ref{fig:fig_coex_curves}, the temperature profiles intersect the miscibility curves at 1--5 GPa, placing them above the EMB. Miscibility temperatures above 10 GPa lie in the solid phase of MgSiO$_3$ as shown in the right panel of Figure~\ref{fig:fig_coex_curves}, making miscible regions above this pressure physically implausible. Silicates and hydrogen are indeed miscible above these pressures. This, however, means that a silicate rain region does not form at these pressures. If silicates and hydrogen are already homogeneously mixed above these pressures, they will remain so \citep{StixrudeGilmore2025, Gupta2025b}. This feature of the miscibility temperatures leads to a different structure compared to the smooth transition proposed by \cite{Rogers2025} in their models (See illustrations in their Figures 2 and 3). Our models here instead suggest a partitioned convective envelope structure, in which the region between the silicate rain layer and the EMB could be convective, and the silicate rain region lies between the EMB and the upper layers. Since there is less mass above the silicate rain region than below it (because it is higher in the envelope), the silicate that is rained to the interior regions generates an inner convective layer between the rain region and the EMB. This creates a bifurcated envelope characterized by an H-He-rich upper envelope and a silicate-rich inner envelope, as illustrated in Figure~\ref{fig:fig_misc_drawing}. Nevertheless, a stably-stratified region bridging the envelope and mantle (e.g., Figure~\ref{fig:fig_homog_vs_inhomog}) and a thinner silicate region located in the outer envelope cause a similar difference in radii. 

The depletion rate depends on the transition layer scale height (100 km used here) and, indirectly, on the equilibrium temperatures and the convective luminosity in the surrounding zones of the rain region. Rain regions that are much colder than the miscibility temperature will deplete more material due to the $Z - Z_{\rm low}$ factor in Eq.~\ref{eq:misc}. Higher temperatures also imply a higher convective luminosity due to higher internal temperatures. Since these models include the effects of convective mixing, which will homogenize compositional gradients (Eq.~\ref{eq:conv_mixing}), convective mixing is a competing process to advection-diffusion (Eq.~\ref{eq:misc}).

\subsection{Implications for Hydrogen--Water Phase Separation}

As Figure~\ref{fig:fig_3_10_evol} illustrates, the envelopes of our sub-Neptune models generally remain above the hydrogen-water miscibility curve of \citet{Gupta2025}, even across a range of equilibrium temperatures, as shown in Figure~\ref{fig:fig_3_teq}. Envelopes with lower metal abundances cool sufficiently to allow water-hydrogen immiscibility at pressures below approximately 30--50 bars, suggesting less water to deplete in the first place. We attempted to deplete water using the advection-diffusion scheme described in Eq.~\ref{eq:misc}. However, the effects of convective mixing (Eq.~\ref{eq:conv_mixing}) dominate the advection-diffusion composition fluxes, so we obtained only modest water depletion in the outermost regions.  We note here that we do not model the effects of condensation in either water or silicate mixtures. Condensation at low pressures and temperatures may occur in these outer regions, separate from immiscibility, which occurs at higher pressures. Consequently, water-poor atmospheres could arise from stable stratification, as demonstrated in Section~\ref{subsec:stratification}, or from condensation if the temperatures are sufficiently cool.

Similar challenges exist in modeling Solar System gas giants. Some ab initio hydrogen–helium miscibility curves are too cold to permit the observed helium rain \citep[e.g.,][]{Schottler2018}. To accurately reproduce the measured helium depletion in Jupiter and Saturn \citep{VonZahn1998b,Koskinen2018}, evolution models typically require shifting immiscibility temperatures to induce helium rain \citep{Nettelmann2015,Pustow2016,Mankovich2020,Tejada2025,Sur2025a}. Analogous adjustments to hydrogen–water miscibility temperatures and their resulting impact on sub-Neptune evolution will be a critical subject for future exploration. Alternatively, different atmospheric models could allow the envelopes to cool much more than those used here, and other heat-transfer assumptions could enhance water depletion, such as manually shutting off convective mixing in the water rain layers. We find no compelling physical reason to manually shut off convective mixing in water rain regions, so we refrain from doing so here.

\subsection{Caveats \& Future Work}\label{subsec:caveats}

The EMBs in all the models presented in Section~\ref{sec:evolution} are hotter than the silicate miscibility curves, implying that the mantles are themselves miscible. We assume that all planets in this work are born with a pure-metal mantle and core. By construction, the mantles do not compositionally diffuse into the envelope, and vice versa, and so hydrogen, water, and other constituents do not diffuse into the mantle. This represents a limitation in our current models, as the diffusion of hydrogen into the mantle may further decrease mantle density and further increase the radii of sub-Neptunes and Earth-sized exoplanets \citep[][]{Chachan2018, Schlichting2022, Gupta2025b} and even for the Earth \citep{Luo2025b}. We will present updated sub-Neptune interior evolution models that allow for metal-hydrogen mixing at the EMB in future work. 

The slower cooling associated with higher equilibrium temperatures, combined with higher convective fluxes, explains why the 800 K equilibrium-temperature model in Figure~\ref{fig:fig_misc_3mearth_50Kvs800K} retains silicates in its outer layer. While this lack of depletion arises from the competition between convective mixing and advection in our simulations, silicate condensation is likely to further deplete the outer layers (an effect not modeled here). Physically, silicates would condense and sink into deeper layers due to the colder outer-envelope temperatures. This phase change could allow complete depletion of silicates from the atmosphere, except perhaps for suspended dust grains, even in the presence of vigorous convection. Additionally, we do not model the mutual phase separation of water and silicates. As both species may be abundant near the potential EMB, this process may be energetically significant and warrants further investigation. 

We model the silicate rain layer with a characteristic scale height, $\mathcal{H}_r$, of 100 km. This is a free parameter in our models; larger values yield wider rain regions but lower depletion rates. A better understanding of silicate droplets in hydrogen-helium mixtures is required to model the scale height and advection rate more accurately. The exact depletion rates depend on the rain scale height, on the instellation, and thermal transport properties of the envelope. This indicates that the precise depletion rate depends on the envelope's thermal and mixing properties, not just on the miscibility curves and equilibrium abundances, which introduces additional uncertainty. Despite these modeling uncertainties, since most of the depletion may occur at early ages, sub-Neptune envelopes are likely to be depleted of silicates by $\gtrsim$ 500 Myr ages. They could hence appear inflated by $\sim$10\% compared to traditional adiabatic models.

There are currently no constraints on the adequate size of the silicate rain layers in sub-Neptune envelopes, as this depends on droplet sizes, diffusion coefficients, and the physics of turbulent diffusion. We follow in the footsteps of past work on the gas giants. While helium rain models rely on calculated droplet sizes to determine sedimentation timescales \citep{Pustow2016, Mankovich2016, Mankovich2020}, the geometry of these layers remains constrained. Estimates for Jupiter’s helium rain region range from a thin layer based on latent heat flux \citep{Markham2024a} to 10\% of the radius based on \textit{Juno} \citep{Bolton2017a} magnetic field data \citep{Wulff2025}. Guided by prior work on helium rain, we adopted a relatively thin, constant silicate rain scale height of 100 km. Increasing this parameter expands the layer but reduces depletion, since the rain layer is spread over a wider area and advection is weaker. Unlike previous helium approaches, we drive silicate depletion via advection proportional to the local metal-abundance excess ($Z - Z_{\rm low}$), rather than to assumed droplet sizes. This mechanism maximizes depletion rates at early ages when $Z - Z_{\rm low}$ is larger. 

Hot sub-Neptunes might appear especially vulnerable to atmospheric escape via photoevaporation \citep[e.g.,][]{Owen2013,ChenRogers2016,OwenWu2017}, core-powered mass loss \citep[e.g.,][]{Ginzburg2018a,Gupta2019,Gupta2022a}, or early boil-off mass loss \citep{Owen2016,Ginzburg2016}. However, recent calculations by \citet{Tang2025b} that couple photoevaporation and boil-off to the thermal evolution of H–He envelopes show that mass loss predominantly affects highly irradiated (F $\gtrsim 100$ F$_{\oplus}$) and low-mass ($\lesssim5$ M$_{\oplus}$) planets when the envelope metallicity is $\gtrsim$50 times solar. We analyzed time-integrated mass loss due to the combined effects of boil-off and photoevaporation for the example models in Sections~\ref{subsec:entropy}, \ref{subsec:stratification}, and \ref{subsec:miscibility}. Core-powered mass loss following boil-off is neglected, as a recent theoretical reassessment by \citet{Tang2024b} found it to be negligible on long evolutionary timescales. We adopt mass-loss prescriptions developed in Section 2.5, Eqs. 13 and 14 of \cite{Tang2025b}. Their coupled hydrodynamics-evolution framework shows that sub-Neptunes may commonly undergo a thermal-energy-mediated phase, in which incident XUV energy is primarily converted into heat and kinetic energy rather than into work against gravity. This post facto mass-loss analysis shows that none of our models are subject to boil-off, and the hot-mantle models in Figure~\ref{fig:fig_3_10_evol} experience only mild photoevaporative loss.\footnote{All the initial conditions of our models should be considered post-formation (i.e., after disk dispersal). } For instance, the 3 M$_\oplus$ model in Figure~\ref{fig:fig_3_10_evol} loses only $\sim 1\%$ of its total mass to photoevaporation over 10 Gyr, compared to an envelope mass fraction of $5\%$, leaving most of the insulating envelope intact. Similar estimates for the stably stratified models in Figure~\ref{fig:fig_homog_vs_inhomog} indicate $\sim$1\% total mass loss. The integrated mass loss for our example models of GJ 1214 b, K2-18 b, TOI-270 d, and TOI-1801 b is $<0.01\%$. Thus, across the parameter space explored here, atmospheric escape influences, but does not determine, the thermal evolution.

In Sections~\ref{subsec:stratification} and \ref{subsec:miscibility}, we further showed that stabilizing silicate gradients associated with mantle–envelope mixing can increase sub-Neptune radii by an additional $\sim$4--7\% relative to homogeneous models. To explore this effect in Section~\ref{subsec:stratification}, we constructed structures in which liquid MgSiO$_3$ mixes with hydrogen and helium above the miscibility curve of \citet{StixrudeGilmore2025} as done by \cite{Rogers2025}. In our implementation, heavy elements are represented by liquid MgSiO$_3$ above the miscibility line (red in Figure~\ref{fig:fig_homog_vs_inhomog}) and by water below it. In reality, MgSiO$_3$ is expected to dissociate into species such as SiO$_2$, MgO, and O$_2$, and to form additional molecules when reacting with water and hydrogen \citep{Schlichting2022}; water itself may also become partially miscible with the mantle, further depleting the envelope and atmosphere \citep{Werlen2025a}. Indeed, recent experiments show that water production from hydrogen-silicate mixing could be prevalent during sub-Neptune formation \citep{Miozzi2025}. Capturing this full chemical complexity with a self-consistent silicate equation of state lies beyond the scope of this work and, more generally, of current evolution models. This is an important direction for future improvements of evolution models in general.

\section{Conclusion}\label{sec:conclusion}

Our sub-Neptune evolution methodology is novel with respect to previous work in the following ways:

\begin{enumerate}
    \item The application and upgrade of \texttt{APPLE}, inspired by stellar evolution codes, and designed to handle self-consistently radiative, conductive, and convective fluxes throughout the entire planetary interior. 
    \item The deployment of an advection diffusion framework to model the evolutionary effects of silicate rain in sub-Neptune envelopes. This advection-diffusion framework drives silicate abundances towards their equilibrium values, informed by the coexistence curves. 
    \item The use of an updated, ab initio liquid MgSiO$_3$ EOS in the mantle \citep{Luo2025}.
    \item The use of an iron alloy EOS, Fe$_{16}$Si \citep{Fischer2012} in the core that aligns more closely to the core characteristics of the Earth's liquid core.
    \item The inclusion of an atmosphere boundary condition that accounts for instellation and atmospheric metallicity \citep{Fortney2020, Ohno2023}.
    \item The inclusion of viscous convection in the partially melted regions of the mantle via modified MLT that follows the rocky planet/super-Earth evolution code developed by \cite{Zhang2022}.
    \item The modeling of the latent heat during the solidification and partial melting regions of the mantle and core, and time-dependent radiogenic heating as implemented in \cite{Zhang2022} in the context of super-Earth interior evolution. 
    \item Applications to model the evolution of four exoplanets: GJ 1214 b, K2-18 b, TOI-270 d, and TOI-1801 b.
\end{enumerate}

Our general conclusions are as follows:

\begin{enumerate}
    \item A hot, liquid rock mantle and core can keep most of its primordial heat over evolutionary timescales and exhibit lower densities than previously expected. As such, sub-Neptune-sized exoplanets with moderate average densities (2--3.5 g cm$^{-3}$) can not be confidently described as water worlds. 
    
    \item An initially stably-stratified composition layer situated immediately above the EMB can further increase the radii of sub-Neptunes across their evolution. This effect compounds with the inefficient cooling caused by the EMB, widening the region of inefficient heat transport and increasing the radius by $\sim$10--15\%, depending on the total metal content and the shape of the stably-stratified layer. 

    \item Silicate rain keeps radii larger by an additional $\sim$5\%, depending on the mass and amount of silicates being depleted. As a result, young, homogeneously mixed sub-Neptunes may show silicate abundances comparable to those of their stars. In contrast, older ($\gtrsim$ 100 Myr) sub-Neptunes will appear depleted and have larger radii than predicted by traditional adiabatic models.

    \item A silicate rain region may not bridge the mantle and the envelope across the EMB if the envelope is initially homogeneous. Instead, the silicate region could be located within the envelope, partitioning it into a silicate-rich inner envelope and an H-He-rich outer envelope. 
    
    \item The radii and average densities of the sub-Neptune exoplanets GJ 1214 b, K2-18 b, TOI-270 d, and TOI-1801 b can be explained by a hot liquid mantle that comprises $\sim$90--95\% of their total mass.

\end{enumerate}

Sub-Neptunes could retain memories of their hot post-formation stages throughout their evolution, allowing further constraints on their initial thermal and compositional states. Using evolutionary models, observations of young planets with \textit{JWST} \citep{Gardner2006} and future missions such as the \textit{Habitable Worlds Observatory} could help constrain the post-formation thermal and compositional states of sub-Neptune exoplanets, thereby informing formation models more effectively. Sub-Neptunes could enhance the memory retention of these post-formation thermal states by creating more stable regions through silicate phase separation. An improved understanding of the microphysical properties, equations of state, modeling methods, and observations of young sub-Neptune exoplanets will increasingly aid our general understanding of planetary formation, interior structure, and exoplanet evolution.

\medskip

This research was funded by the Center for Matter at Atomic Pressures (CMAP), a National Science Foundation (NSF) Physics Frontier Center under Award PHY-2020249. Any opinions, findings, conclusions, or recommendations expressed herein are those of the authors and do not necessarily reflect NSF views. RTA is grateful to Drs. Ankan Sur and Yubo Su for lively discussions, critical feedback, expertise, and academic mentorship. RTA is thankful to Dr. Jisheng Zhang for providing his evolution code, which guided upgrades to the viscous convection in our own evolution code. RTA is grateful to Drs. Kazumasa Ohno and Jonathan Fortney for providing their atmosphere models, and to Dr. Yihang Peng for guidance on his conductivities. AG acknowledges support from the Heising–Simons Foundation through the 51 Pegasi b Fellowship, and from Princeton University through the Harry H. Hess Fellowship and the Future Faculty in Physical Sciences Fellowship. The calculations presented in this article were performed on computational resources managed and supported by Princeton Research Computing, a consortium comprising the Princeton Institute for Computational Science and Engineering (PICSciE) and the Office of Information Technology's High Performance Computing Center and Visualization Laboratory at Princeton University. RTA thanks Dr. Matthew Coleman from Princeton Research Computing for computational and technical assistance.



\begin{deluxetable*}{lccccc}[t!]
\tablecaption{Properties of selected sub-Neptune exoplanets\label{tab:subneps}}
\tablehead{
\colhead{Planet} &
\colhead{$M_p$ [$M_\oplus$]} &
\colhead{$R_p$ [$R_\oplus$]} &
\colhead{$T_{\rm eq}$ [K]} &
\colhead{Estimated Age [Gyr]} &
\colhead{Mean Densities [g cm$^{-3}$]}
}
\startdata
GJ 1214 b & $8.17 \pm 0.43$ & $2.74^{+0.050}_{-0.053}$ & $596 \pm 19$ & 6--10 & 2.2$^{+0.17}_{-0.16}$\\
K2-18 b & $8.63 \pm 1.35$ & $2.61 \pm 0.09$ & $265 \pm 5$ & $2.4\pm0.4$ & 2.67$^{+0.52}_{-0.47}$\\
TOI-270 d & $4.78 \pm 0.43$ & $2.133 \pm 0.058$ & $354 \pm 8$ & 1--10 & $2.72\pm0.33$\\
TOI-1801 b & $5.74 \pm 1.46$ & $2.08 \pm 0.12$ & $\sim$440 & 0.6--0.8 & $3.7 \pm 1.22$ 
\enddata
\tablecomments{
Masses and radii are taken from the references listed in the final column.
For GJ~1214~b, we list the mass and radius from \citet{Cloutier2021}; the system's age is estimated to be $6$--$10$~Gyr from star spot rotation \citep{Mallonn2018}.
For K2-18 b, we adopt the mass and radius from \cite{Cloutier2019, Benneke2019} and the age from \citet{Guinan2019}.
We adopt the TOI-270 d values reported in \cite{VanEylen2021}. We note that the ages of TOI-270 itself are not yet constrained.
For TOI-1801 b, we use the mass, radius, and age reported by \citet{Mallorquin2023}.
The equilibrium temperature of $\sim 600$ K of GJ 1214 b was taken from \cite{Gao2023a};
for K2-18 b we list $T_{\rm eq}=265\pm5$~K from stellar and orbital parameters \citep[e.g.,][]{Cloutier2017};
for TOI-1801 b we adopt the reported value $T_{\rm eq}\approx440$~K \citep{Mallorquin2023}
}
\end{deluxetable*}

\clearpage

\bibliography{references}{}
\bibliographystyle{aasjournal}

\end{document}